\documentclass[usegraphicx,usenatbib,fleqn,a4paper]{mn2e}
\usepackage{color}
\usepackage{times}

\newcommand{\kms}{\ensuremath{\mathrm{km~s}^{-1}}}

\newcommand{\Msol}{\ensuremath{M_\odot}}
\newcommand*{\rom}[1]{\romannumeral #1}
\newcommand{\ion}[2]{{#1}~\textsc{\rom{#2}}}

\newcommand{\etal}{\emph{et al.}}
\newcommand{\aj}{AJ}
\newcommand{\aanda}{A\&A}
\newcommand{\apj}{ApJ}

\newcommand{\apjs}{ApJS}
\newcommand{\astroph}[1]{arXiv:astro-ph/{#1}}
\newcommand{\arxiv}[1]{arXiv:{#1}}
\newcommand{\mnras}{MNRAS}
\newcommand{\pasp}{PASP}
\newcommand{\prd}{Phys. Rev. D}
\newcommand{\prl}{Phys. Rev. Lett.}

\newcommand{\na}{\ion{Na}{1}~D}
\newcommand{\ewna}{\ensuremath{EW(\mathrm{\ion{Na}{1}~D)}}}
\newcommand{\nickel}{{\ensuremath{^{56}\mathrm{Ni}}}}
\newcommand{\cobalt}{{\ensuremath{^{56}\mathrm{Co}}}}

\newcommand{\lcdm}{\ensuremath{\Lambda}CDM}
\newcommand{\MNi}{\ensuremath{M_{^{56}\mathrm{Ni}}}}
\newcommand{\MFe}{\ensuremath{M_\mathrm{Fe}}}
\newcommand{\MWD}{\ensuremath{M_\mathrm{ej}}}
\newcommand{\snf}[1]{\mbox{SNF~20{#1}}}

\newcommand{\vKE}{\ensuremath{v_\mathrm{KE}}}
\newcommand{\aNi}{\ensuremath{a_\mathrm{Ni}}}
\newcommand{\nb}[1]{\ensuremath{^\mathrm{#1}}}

\newcommand{\revised}[1]{{\textcolor{black} {#1}}}

\begin{document}


\title[SN I\lowercase{a} Ejected Masses from SNfactory]
      {Type I\lowercase{a} supernova bolometric light curves
       and ejected mass estimates
       from the Nearby Supernova Factory}

\author[Scalzo et al.]
{
    R.~Scalzo$^{1,2,}$\thanks{Email: richard.scalzo@anu.edu.au},
    G.~Aldering$^3$,
    P.~Antilogus$^4$,
    C.~Aragon$^{3,5}$,
    S.~Bailey$^3$,
    C.~Baltay$^6$, \newauthor
    S.~Bongard$^4$,
    C.~Buton$^{7,8}$,
    F.~Cellier-Holzem$^4$,
    M.~Childress$^{1,2}$,
    N.~Chotard$^9$,
    Y.~Copin$^9$, \newauthor
    H.~K. Fakhouri$^{3,10}$,
    E.~Gangler$^9$,
    J.~Guy$^4$,
    A.~Kim$^3$,
    M.~Kowalski$^7$,
    M.~Kromer$^{11}$, \newauthor
    J.~Nordin$^3$,
    P.~Nugent$^{12,13}$,
    K.~Paech$^7$,
    R.~Pain$^{2,4}$,
    E.~Pecontal$^{14}$,
    R.~Pereira$^9$, \newauthor
    S.~Perlmutter$^{3,10}$,
    D.~Rabinowitz$^6$,
    M.~Rigault$^9$,
    K.~Runge$^3$,
    C.~Saunders$^3$,
    S.~A.~Sim$^{1,2,15}$, \newauthor
    G.~Smadja$^9$,
    C.~Tao$^{16,17}$,
    S.~Taubenberger$^{11}$,
    R.~C. Thomas$^{12}$,
    and B.~A.~Weaver$^{18}$ \newauthor
    \mbox{(The~Nearby~Supernova~Factory)} \\
    $^{ 1}$ Research School of Astronomy and Astrophysics,
            The Australian National University,
            Cotter Road, Weston Creek, ACT 2611, Australia \\
    \revised{
    $^{ 2}$ ARC Centre of Excellence for All-Sky Astrophysics (CAASTRO)} \\
    $^{ 3}$ Physics Division, Lawrence Berkeley National Laboratory, 
            1 Cyclotron Road, Berkeley, CA 94720, USA \\
    $^{ 4}$ Laboratoire de Physique Nucl\'eaire et des Hautes \'Energies,
            Universit\'e Pierre et Marie Curie Paris 6,
            Universit\'e Paris Diderot \\ Paris 7, CNRS-IN2P3, 
            4 place Jussieu, 75252 Paris Cedex 05, France \\
    $^{ 5}$ Present address:
            Department of Human Centered Design \& Engineering,
            University of Washington,
            423 Sieg Hall, Box 352315, Seattle, WA 98195, USA \\
    $^{ 6}$ Department of Physics, Yale University, 
            New Haven, CT, 06250-8121, USA \\
    $^{ 7}$ Physikalisches Institut, Universit\"at Bonn,
            Nu\ss allee 12, 53115 Bonn, Germany \\
    $^{ 8}$ Present address:
            Synchrotron Soleil, L'Orme des Merisiers,
            Saint-Aubin — BP 48 91192, GIF-sur-YVETTE CEDEX, France \\
    $^{ 9}$ Universit\'e de Lyon, F-69622, Lyon, France;
            Universit\'e de Lyon 1, Villeurbanne; 
            CNRS/IN2P3, Institut de Physique Nucl\'eaire de Lyon \\
    $^{10}$ Department of Physics, University of California, Berkeley,
            366 LeConte Hall MC 7300, Berkeley, CA 94720-7300, USA \\
    $^{11}$ Max-Planck-Institut f¨ur Astrophysik, Karl-Schwarzschild-Str. 1,
            85741 Garching bei M\"unchen, Germany \\
    $^{12}$ Computational Cosmology Center, Computational Research Division,
            Lawrence Berkeley National Laboratory, 
            1 Cyclotron Road \\ MS 50B-4206, Berkeley, CA 94720, USA \\
    $^{13}$ Department of Astronomy, University of California, Berkeley,
            B-20 Hearst Field Annex \#3411, Berkeley, CA 94720-3411, USA \\
    $^{14}$ Centre de Recherche Astronomique de Lyon, Universit\'e Lyon 1,
            9 Avenue Charles Andr\'e, 69561 Saint Genis Laval Cedex, France \\
    $^{15}$ Astrophysics Research Centre, School of Mathematics and Physics,
            Queen's University Belfast, Belfast BT7 1NN, UK \\
    $^{16}$ Centre de Physique des Particules de Marseille, 163,
            avenue de Luminy - Case 902 - 13288 Marseille Cedex 09, France \\
    $^{17}$ Tsinghua Center for Astrophysics, Tsinghua University,
            Beijing 100084, China \\
    $^{18}$ Center for Cosmology and Particle Physics, New York University,
            4 Washington Place, New York, NY 10003, USA \\
}

\maketitle

\begin{abstract}
We present a sample of normal type~Ia supernovae from the Nearby Supernova
Factory dataset with spectrophotometry at sufficiently late phases to
estimate the ejected mass using the bolometric light curve.
We measure \nickel\ masses from the peak bolometric luminosity, then compare
the luminosity in the \cobalt-decay tail to the expected rate of radioactive
energy release from ejecta of a given mass.  We infer the ejected mass in a
Bayesian context using a semi-analytic model of the ejecta, incorporating
constraints from contemporary numerical models as priors on the density
structure and distribution of \nickel\ throughout the ejecta.
We find a strong correlation between ejected mass and light curve
decline rate, and consequently \nickel\ mass, with ejected
masses in our data ranging from 0.9--1.4~\Msol.
Most fast-declining \mbox{(SALT2 $x_1 < -1$)} normal SNe~Ia have significantly
sub-Chandrasekhar ejected masses in our fiducial analysis.
\end{abstract}



\section{Introduction}

Type Ia supernovae (SNe~Ia) have been used for well over a decade as precision
luminosity distance indicators, leading to the discovery of the universe's
accelerated expansion \citep{riess98,scp99} \revised{which has been measured
in contemporary studies with increasing precision}
\citep{hicken09,kessler09,sullivan11b,suzuki12}.
SN~Ia luminosities can be measured to an accuracy of $\sim 0.15$~mag using
correlations between the luminosity, colour, and light curve width
\citep{riess96,tripp98,phillips99,goldhaber01}, and
many recent and ongoing studies have sought to further reduce this dispersion
by looking for new correlations between SN~Ia luminosities and their
spectroscopic properties \citep{sjb09,wang09,csp10,fk10}.

The spectra of SNe~Ia show no hydrogen, no helium, and strong
intermediate-mass element signatures; they are generally
understood to be thermonuclear explosions of carbon/oxygen white dwarfs
in binary systems.  The absence of a detectable shock breakout in the early
light curve of the nearby SN~Ia~2011fe \citep{nugent11,bloom12} provides
direct evidence that the progenitor primary must be a compact object such as
a white dwarf.  However, many variables remain which can affect the explosion,
including the evolutionary state of the white dwarf progenitor's binary
companion, the circumstellar environment, the explosion trigger,
and the progress of nuclear burning in the explosion.
The low luminosities, small radii, and relatively clean environments of
white dwarfs make SN~Ia progenitor systems notoriously hard to constrain.
Uncovering the nature of SN~Ia progenitor systems and explosions is therefore
an interesting puzzle in its own right.  From a cosmological viewpoint, if
two or more SN~Ia progenitor channels exist which have slightly different
peak luminosities or luminosity standardization relations, and their relative
rates evolve with redshift, the resulting shift in the mean luminosity could
mimic a time-varying dark energy equation of state \citep{linder06}.

The two main competing SN~Ia progenitor scenarios are the
\emph{single-degenerate} scenario \citep{wi73}, in which a carbon/oxygen
white dwarf slowly accretes mass from a non-degenerate companion until exploding
near the Chandrasekhar mass, and the \emph{double-degenerate} scenario
\citep{it84}, in which two white dwarfs collide or merge.  The classical
formulations of these scenarios assume the primary white dwarf must explode
near the Chandrasekhar limit; however, in the
\emph{sub-Chandrasekhar double-detonation} variant,
a sub-Chandrasekhar-mass white dwarf can be made to explode by
the detonation of a layer of helium on its surface, accreted from the binary
companion \citep{ww94,sim10,fink10,kromer10,sim12}.
Distinguishing which of these models accounts for the majority of
spectroscopically ``normal'' \citep{bfn93}, hence cosmologically useful,
SNe~Ia has been a very active subject of current research
\citep[for a recent review see][]{wh12}.
Binary population synthesis models of the Chandrasekhar-mass single-degenerate
and double-degenerate channels often have trouble producing enough SNe~Ia to
reproduce the observed rate \revised{\citep[but see][]{hp04,ruiter11}};
this is one of the main motivations for investigating
sub-Chandrasekhar models \citep{vkcj10}.

The mass of the progenitor is a fundamental physical variable with
power to differentiate between different progenitor scenarios.
While Chandrasekhar-mass delayed detonations have been historically favored,
viable super-Chandrasekhar-mass evolution pathways and explosion models have
been proposed for both single-degenerate \citep{justham11,hachisu11,rds12}
and double-degenerate \citep{pakmor10,pakmor11,pakmor12} SN~Ia progenitors,
and sub-Chandrasekhar-mass models must necessarily involve a different
explosion trigger than any of these.  The white dwarf progenitor is totally
disrupted in theoretical models of normal SNe~Ia, although a bound remnant
may remain in some models which try to reproduce underluminous, peculiar
events such as SN~2002cx \citep{kromer13}.  For normal SNe~Ia, then,
measuring the progenitor mass reduces to measuring the ejected mass.
Nebular-phase spectra can be used to estimate
the mass of iron-peak elements in the ejecta \citep[e.g.][]{mazzali07},
but only the closest SNe~Ia are bright enough to yield high-quality spectra
in nebular phase $\sim 1$~year after explosion, which limits
the number of SNe on which this technique can be used.

\citet{stritz06} used SN~Ia quasi-bolometric light curves ($UBVRI$) in early
nebular phase (50--100~days after $B$-band maximum light) to estimate the
ejected mass, as follows:  The mass of \nickel,
\revised{the radioactive decay of which} powers the near-maximum
light curve of normal SNe~Ia, can be inferred from the bolometric luminosity
at maximum light \citep{arnett82}.  The decay of \cobalt, itself a decay
product of \nickel, powers the post-maximum light curve.
At sufficiently late times, the shape of the bolometric light curve is
sensitive to the degree of trapping of gamma rays from \cobalt\ decay
\citep{jeffery99}; greater ejected masses provide greater optical depth to
Compton scattering, and hence higher luminosity, for a given phase and
\nickel\ mass.
\citet{scalzo10,scalzo12} refined this method by including more accurate
near-infrared (NIR) corrections and a set of prior constraints on model
inputs from contemporary explosion models, using it to estimate the masses
of several candidate super-Chandrasekhar-mass SNe~Ia; they found ejected
masses of $2.30^{+0.27}_{-0.24}$~\Msol\ for the superluminous SN~Ia~2007if
and $1.79^{+0.28}_{-0.21}$~\Msol\ for the spectroscopically 1991T-like
\snf{080723-012}, interpreting them as double-degenerate explosions powered
entirely by radioactive decay.

In the current work, we use this method as implemented in \citet{scalzo12}
on a set of \emph{normal} SNe~Ia, attempting to quantify the distribution
of progenitor mass scales in the context of different progenitor scenarios.
Our supernova discoveries, our sample selection, and the provenance of
our data are described in \S\ref{sec:observations}.
Our \revised{method for constructing} full $UBVRIYJHK$
(3300--\revised{23900}~\AA)
bolometric light curves for 19 spectroscopically normal SNe~Ia,
(including NIR corrections for the $YJHK$ flux which we do not observe),
are presented in \S\ref{sec:analysis}.  We briefly review the assumptions
of our ejected mass reconstruction method in \S\ref{sec:modeling},
and present the reconstructed masses for our 19 SNe.  We also present
\revised{ejected mass and \nickel\ mass reconstructions based on}
synthetic observables from a series of contemporary explosion models.
In \S\ref{sec:discussion} we examine correlations between
\revised{ejected mass} and other quantities, such as photospheric light
curve fit parameters (decline rate and colour) and \nickel\ \revised{mass}.
We summarize and conclude in \S\ref{sec:conclusions}.


\section{Observations}
\label{sec:observations}

All supernova observations in this paper
were obtained with the SuperNova Integral Field Spectrograph
\citep[SNIFS;][]{snf,snifs}, built and operated by the SNfactory.
SNIFS is a fully integrated instrument optimized
for automated observation of point sources on a structured background
over the full optical window at moderate spectral resolution.  It
consists of a high-throughput wide-band pure-lenslet integral field
spectrograph \citep[IFS;][]{bacon95,bacon00,bacon01},
a multifilter photometric channel to image the field surrounding the IFS
for atmospheric transmission monitoring simultaneous with spectroscopy,
and an acquisition/guiding channel.  The IFS possesses a fully filled
$6\farcs 4 \times 6\farcs 4$ spectroscopic field of view (FOV) subdivided
into a grid of $15 \times 15$ spatial elements (spaxels), a dual-channel
spectrograph covering 3200--5200~\AA\ and 5100--10000~\AA\ simultaneously,
and an internal calibration unit (continuum and arc lamps).
SNIFS is continuously mounted on the south bent Cassegrain port of the
UH 2.2-meter telescope (Mauna Kea) and is operated remotely.


\subsection{Discovery}

Thirteen of the SNe studied in this paper are among the 400 SNe Ia discovered
in the SNfactory SN Ia search, carried out between 2005 and 2008 with the
QUEST-II camera \citep{baltay07} mounted on the Samuel Oschin 1.2-m Schmidt
telescope at Palomar Observatory (``Palomar/QUEST'').  QUEST-II observations
were taken in a broad RG-610 filter with appreciable transmission from
6100--10000~\AA, covering the Johnson $R$ and $I$ bandpasses.  Upon
discovery\revised{,} candidate SNe were spectroscopically screened using SNIFS.
Our normal criteria for continuing spectrophotometric follow-up of SNe~Ia
with SNIFS were that the spectroscopic phase be at or before maximum light,
as estimated using a template-matching code similar e.g. to SUPERFIT
\citep{superfit}, and that the redshift be in the range $0.03 < z < 0.08$.

We also include six SNe from other searches which have extensive coverage
with SNIFS from maximum light to 40 days or more after maximum light:
PTF09dlc and PTF09dnl \revised{\citep{ptf09dlc}} and SN~2011fe
\citep{sn2011fe}, discovered by the Palomar Transient Factory (PTF);
SN 2005el \citep{sn2005el} and SN 2008ec \citep{sn2008ec}, discovered by
the Lick Observatory Supernova Search (LOSS); and SN 2007cq \citep{sn2007cq},
discovered by T. Orff and J. Newton.


\subsection{Follow-up Observations and Reduction}

The SNIFS spectrophotometric data reduction pipeline has been described
in previous papers \citep{bacon01,snf2005gj,scalzo10,buton13}.  We subtract
the host galaxy light in both spatial directions using the methodology
described in \citet{bongard11}, which uses SNIFS IFS exposures of the host
taken after each SN has faded away.

The photometry used for the modeling in this paper was synthesized from
SNIFS flux-calibrated rest-frame spectra, corrected for Galactic dust
extinction using $E(B-V)$ from \citet{sfd} and the extinction law of
\citet{cardelli} with $R_V = 3.1$.  Redshifts were obtained from host galaxy
spectra as described in \revised{\citet{childress13}}.


\subsection{Sample Selection}
\label{subsec:selection}

The supernovae we chose to study in this paper were selected from the
currently processed sample of 147 SNe~Ia followed spectrophotometrically
with SNIFS, as follows.

\revised{To include a SN in our sample, we require that it be}
spectroscopically typed via SNID
\citep{snid} as ``Ia-norm'', using a spectrum at or before maximum light,
\revised{and that it is not obviously highly reddened.}
This removes the highly reddened \snf{080720-001}, as well as
SN~2007if and spectroscopically 1991T-like events \citep{scalzo10,scalzo12}.
We include the peculiar SNe~Ia from \cite{scalzo12}, as well as a single
1999aa-like event (\snf{070506-006}), in some of our plots for visual
comparison, but exclude them from discussion of the distribution of
properties of normal events.

We also require full 3300--8800~\AA\ wavelength
coverage with SNIFS for epochs near maximum light and at sufficiently late
phase to determine the bolometric luminosity at maximum and at least 40 days
after $B$-band maximum light.  By performing repeated fits of several of our
SNe with different scaling factors for the late-time error bars, we assessed
how the precision and accuracy of the fit depend on the combined precision
of the late-time light curve data points (see \S\ref{subsec:xchckes}).
We found that a total exposure with stacked signal-to-noise greater than 15
(or a single point with error bar less than 0.06~mag) at rest-frame
$B$-band phases past +40~days was required in order to accurately constrain
the ejected mass.  This limit was insensitive to the number or relative
phases of light curve points.  \revised{Above} this target signal-to-noise
our ejected mass estimates are systematics-dominated, mostly by nuisance
parameters over which we marginalize in our analysis; beneath it, our fits
rapidly lose constraining power.
Since SNfactory's main science goal is SN~Ia Hubble diagram cosmology,
which does not require late-time observations except for host galaxy
subtraction, few SNfactory SNe~Ia have light curve coverage at later
phases than about 35 days past $B$-band maximum light.
After this cut, we have 23 SNe remaining.

We cut an additional 3 SNe~Ia for which the flux calibration was too uncertain
due to poor observing conditions during late-time observations, introducing
large systematic fluctuations into their light curves.  We were able to
identify these points by the large residuals of the corresponding SNIFS data
cubes from a model of the host galaxy plus point source at that epoch
produced by the method of \cite{bongard11}.  The quality of these light
curves should improve with planned processing improvements, but we do not
include these SNe in the present sample.

Finally, we remove the very nearby supernova SN~2009ig ($z = 0.0087$),
for which a reasonable assumption for the random peculiar motion of 300~\kms\
leads to a large (0.25~mag) error on the distance modulus, but for which the
only independent distance measurement is a highly uncertain (0.4~mag)
Tully-Fisher distance modulus.  This large uncertainty in
\revised{distance produces a large corresponding uncertainty in luminosity,
and hence \nickel\ mass, which} makes it impossible to
determine the characteristics of SN~2009ig with reasonable precision.
Our final sample therefore contains 19 SNe~Ia.


\section{Analysis}
\label{sec:analysis}

In this section we discuss the construction of bolometric light curves from
SNfactory spectrophotometry.  We use Gaussian process regression
extensively as a convenient interpolation technique for our data, which we
describe in more detail in Appendix~\ref{sec:gp}; for a more comprehensive
introduction, see \citet{rw06}.  We describe here how we characterize the
synthetic broad-band light curves of our SNe~Ia and estimate host galaxy
extinction (\S\ref{subsec:hostred}); how we estimate the flux at NIR
wavelengths unobserved by SNIFS in \S\ref{subsec:nircorr}; and how we
integrate the flux density over wavelength and produce final bolometric
light curves in \S\ref{subsec:bololc}.


\subsection{Light Curve Characteristics and Extinction}
\label{subsec:hostred}

\revised{We synthesized} multi-band photometry from SNIFS flux-calibrated
spectra in wavelength regions corresponding approximately to Bessell
$B$, $V$, and $R$
\citep[see][]{sjb09}, and these light curves were fit using SALT2
\citep{guy07,guy10}.  The light curve shape parameter $x_1$ and colour
$c$ are listed in Table~\ref{tbl:snfsalt}.  The SN host galaxy redshifts,
listed in the same table, are from \citet{childress13}.

\begin{table*}
\caption{SALT2 light curve fit inputs and fit results}
\newcommand{\aatag}{\ensuremath{^\dagger}}
\begin{tabular}{lrrcccrrrr}
\hline 
SN~Name & $z_\mathrm{helio}$ & $z_\mathrm{CMB}$ & \revised{$E(B-V)_\mathrm{MW}$}
   & \revised{MJD($B_\mathrm{max}$)} & \revised{$M_{B,\mathrm{max}}^{a}$} & SALT2~$x_1$ & SALT2~$c$ \\ 
        & & & \revised{(mag)} & \revised{(days)} & \revised{(mag)} & & \\ 
\hline 
\multicolumn{8}{c}{SNfactory-Discovered Supernovae} \\
\hline 
SNF~20060907-000 & 0.05731 & 0.05624 & 0.152 & 53993.7 
                 & $-19.44 \pm 0.04$ & $-0.70 \pm 0.18$ & $-0.122 \pm 0.015$ \\ 
SNF~20061020-000 & 0.03841 & 0.03723 & 0.031 & 54035.8
                 & $-18.82 \pm 0.06$ & $-1.74 \pm 0.25$ & $ 0.079 \pm 0.029$ \\ 
SNF~20070506-006\aatag & 0.03491 & 0.03554 & 0.046 & 54243.6
                 & $-19.48 \pm 0.05$ & $ 1.06 \pm 0.14$ & $ 0.049 \pm 0.017$ \\ 
SNF~20070701-005 & 0.06958 & 0.06832 & 0.031 & 54283.6
                 & $-19.43 \pm 0.04$ & $-0.38 \pm 0.14$ & $ 0.082 \pm 0.013$ \\ 
SNF~20070810-004 & 0.08394 & 0.08268 & 0.040 & 54331.2
                 & $-19.17 \pm 0.02$ & $-0.32 \pm 0.12$ & $ 0.056 \pm 0.011$ \\ 
SNF~20070817-003 & 0.06400 & 0.06299 & 0.032 & 54336.9
                 & $-18.95 \pm 0.04$ & $-1.23 \pm 0.16$ & $-0.014 \pm 0.015$ \\ 
SNF~20070902-018 & 0.06908 & 0.06799 & 0.036 & 54351.8
                 & $-18.80 \pm 0.03$ & $-0.85 \pm 0.12$ & $-0.232 \pm 0.033$ \\ 
SNF~20080522-011 & 0.03789 & 0.03846 & 0.043 & 54616.7
                 & $-19.48 \pm 0.05$ & $ 0.69 \pm 0.20$ & $-0.006 \pm 0.016$ \\ 
SNF~20080620-000 & 0.03307 & 0.03332 & 0.067 & 54641.3
                 & $-18.83 \pm 0.06$ & $-1.04 \pm 0.18$ & $ 0.118 \pm 0.018$ \\ 
SNF~20080717-000 & 0.05937 & 0.05817 & 0.053 & 54672.6
                 & $-18.56 \pm 0.03$ & $ 0.87 \pm 0.15$ & $ 0.242 \pm 0.013$ \\ 
SNF~20080803-000 & 0.05706 & 0.05706 & 0.073 & 54690.5
                 & $-18.82 \pm 0.04$ & $ 0.26 \pm 0.15$ & $ 0.200 \pm 0.014$ \\ 
SNF~20080913-031 & 0.05485 & 0.05395 & 0.081 & 54732.5
                 & $-19.12 \pm 0.04$ & $-0.14 \pm 0.23$ & $ 0.053 \pm 0.016$ \\ 
SNF~20080918-004 & 0.05100 & 0.04990 & 0.042 & 54734.5
                 & $-18.95 \pm 0.05$ & $-1.83 \pm 0.29$ & $-0.021 \pm 0.024$ \\ 
\hline 
\multicolumn{8}{c}{Externally Discovered Supernovae Observed by SNfactory} \\
\hline 
SN2005el         & 0.01491 & 0.01490 & 0.114 & 53646.6
                 & $-19.36 \pm 0.13$ & $-2.20 \pm 0.18$ & $-0.140 \pm 0.031$ \\ 
SN2007cq         & 0.02578 & 0.02456 & 0.110 & 54280.8
                 & $-19.39 \pm 0.08$ & $-0.72 \pm 0.18$ & $ 0.005 \pm 0.019$ \\ 
SN2008ec         & 0.01632 & 0.01507 & 0.069 & 54673.9
                 & $-18.60 \pm 0.13$ & $-1.61 \pm 0.17$ & $ 0.212 \pm 0.023$ \\ 
SN2011fe         & 0.00080 & 0.00080 & 0.009 & 55814.5
                 & $-19.10 \pm 0.12$ & $-0.21 \pm 0.07$ & $-0.066 \pm 0.021$ \\ 
PTF09dlc         & 0.06750 & 0.06628 & 0.054 & 55075.2
                 & $-19.31 \pm 0.03$ & $-0.10 \pm 0.11$ & $-0.007 \pm 0.010$ \\ 
PTF09dnl         & 0.02310 & 0.02297 & 0.043 & 55075.0
                 & $-19.13 \pm 0.09$ & $ 0.62 \pm 0.14$ & $ 0.146 \pm 0.013$ \\ 
\hline 
\end{tabular}
\medskip \\
\flushleft
$^a$~Includes error in distance modulus, measured either from most accurate available independent distance
     or (for Hubble-flow SNe) by using the $\Lambda$CDM luminosity distance
     ($\Omega_\Lambda = 0.72$, $\Omega_K = 0.00$, $H_0 = 72$~km~s$^{-1}$~Mpc$^{-1}$)
     and assuming a 300~km~s$^{-1}$ random peculiar velocity error.\\
\aatag~Typed by SNID as 1999aa-like from multiple pre-maximum spectra.

\label{tbl:snfsalt}
\end{table*}

As in \citet{scalzo12}, we estimate host galaxy extinction in two different
ways.  First, we fit the $B-V$ colour behavior of each SN
to the Lira relation \citep{phillips99,csp10}, since we have at least one
observation later than $B$-band phase +30 days for each SN.
Additionally, we search for \na\ absorption at the redshift of
the host galaxy for each SN.  We perform a $\chi^2$ fit to the \na\ line
profile, modeled as two separate Gaussian lines with full width at half
maximum equal to the SNIFS instrumental resolution of 6~\AA, to all SNIFS
spectra of each SN.  In the fit, the equivalent width \ewna\
of the \na\ line is constrained to be non-negative.
We convert these to estimates of $E(B-V)_\mathrm{host}$ using the relation
of \citet{ppb12}, which we find corresponds roughly to the shallow-slope
(0.16~mag~\AA$^{-1}$) relation of \citet{tbc02} for low equivalent width,
but which produces less tension with the Lira relation and the fitted
SALT2 colours of our SNe for $\ewna > 1.0$~\AA.
To increase the precision of our final reddening estimates,
we combine information about host galaxy extinction from \ewna\ and from
the Lira relation.
The best-fitting Lira excesses, values of \ewna, and final derived constraints
on the host galaxy reddening are listed in Table~\ref{tbl:hostred}.

\begin{table*}
\caption{Host reddening measures}
\newcommand{\ctr}[1]{\multicolumn{1}{c}{\revised{#1}}}
\begin{tabular}{lrrrr}
\hline 
SN~Name & \ctr{\ewna$^a$} & \ctr{$E(B-V)_\mathrm{\na}^a$} & \ctr{$E(B-V)_\mathrm{Lira}^b$} & \ctr{$E(B-V)_\mathrm{joint}$} \\
        & \ctr{(\AA)}     & \ctr{(mag)}                   & \ctr{(mag)}                    & \ctr{(mag)} \\
\hline 
\multicolumn{5}{c}{SNfactory-Discovered Supernovae} \\
\hline 
SNF20060907-000 & $<               0.23$ & $<               0.03$ & $-0.51 \pm 0.08$ & $0.02^{+0.01}_{-0.01}$ \\[0.2ex]
SNF20061020-000 & $<               0.34$ & $<               0.04$ & $-0.01 \pm 0.08$ & $0.02^{+0.01}_{-0.01}$ \\[0.2ex]
SNF20070506-006 & $<               0.17$ & $<               0.03$ & $ 0.09 \pm 0.08$ & $0.01^{+0.01}_{-0.01}$ \\[0.2ex]
SNF20070701-005 & $0.70^{+0.18}_{-0.19}$ & $0.09^{+0.07}_{-0.04}$ & $ 0.07 \pm 0.08$ & $0.08^{+0.04}_{-0.03}$ \\[0.2ex]
SNF20070810-004 & $<               0.11$ & $<               0.02$ & $ 0.14 \pm 0.08$ & $0.00^{+0.01}_{-0.00}$ \\[0.2ex]
SNF20070817-003 & $<               0.30$ & $<               0.03$ & $-0.13 \pm 0.08$ & $0.01^{+0.01}_{-0.01}$ \\[0.2ex]
SNF20070902-018 & $0.77^{+0.24}_{-0.27}$ & $0.11^{+0.11}_{-0.06}$ & $-0.14 \pm 0.08$ & $0.04^{+0.03}_{-0.02}$ \\[0.2ex]
SNF20080522-011 & $<               0.11$ & $<               0.02$ & $ 0.03 \pm 0.08$ & $0.00^{+0.01}_{-0.00}$ \\[0.2ex]
SNF20080620-000 & $<               0.19$ & $<               0.03$ & $-0.05 \pm 0.10$ & $0.01^{+0.01}_{-0.01}$ \\[0.2ex]
SNF20080717-000 & $1.13^{+0.14}_{-0.15}$ & $0.30^{+0.17}_{-0.11}$ & $ 0.27 \pm 0.08$ & $0.26^{+0.07}_{-0.08}$ \\[0.2ex]
SNF20080803-000 & $0.92^{+0.13}_{-0.14}$ & $0.17^{+0.08}_{-0.06}$ & $ 0.12 \pm 0.09$ & $0.15^{+0.05}_{-0.04}$ \\[0.2ex]
SNF20080913-031 & $<               0.17$ & $<               0.02$ & $-0.01 \pm 0.08$ & $0.00^{+0.01}_{-0.00}$ \\[0.2ex]
SNF20080918-004 & $<               0.16$ & $<               0.02$ & $ 0.03 \pm 0.08$ & $0.00^{+0.01}_{-0.00}$ \\[0.2ex]
\hline 
\multicolumn{5}{c}{Externally Discovered Supernovae Observed by SNfactory} \\
\hline 
SN~2005el        & $0.11^{+0.03}_{-0.03}$ & $0.02^{+0.01}_{-0.01}$ & $-0.10 \pm 0.08$ & $0.02^{+0.01}_{-0.01}$ \\[0.2ex]
SN~2007cq        & $<               0.08$ & $<               0.02$ & $ 0.03 \pm 0.08$ & $0.00^{+0.01}_{-0.00}$ \\[0.2ex]
SN~2008ec        & $0.57^{+0.03}_{-0.03}$ & $0.06^{+0.01}_{-0.01}$ & $ 0.18 \pm 0.08$ & $0.07^{+0.02}_{-0.01}$ \\[0.2ex]
SN~2011fe        & $<               0.16$ & $<               0.02$ & $ 0.00 \pm 0.08$ & $0.00^{+0.00}_{-0.00}$ \\[0.2ex]
PTF09dlc         & $<               0.15$ & $<               0.02$ & $-0.03 \pm 0.08$ & $0.00^{+0.01}_{-0.00}$ \\[0.2ex]
PTF09dnl         & $0.10^{+0.02}_{-0.02}$ & $0.02^{+0.01}_{-0.01}$ & $ 0.13 \pm 0.08$ & $0.02^{+0.01}_{-0.01}$ \\[0.2ex]
\hline 
\end{tabular}
\medskip \\
\flushleft
$^a$ Listed error bars are 68\% CL (``$1\sigma$'') errors.  When the \na\ line was not detected at greater than
$2\sigma$ (95\% CL), upper limits on \ewna\ and $E(B-V)_{\na}$ are 95\% CL. \\
$^b$ Errors dominated by systematic scatter around the Lira relation \citep{csp10}.

\label{tbl:hostred}
\end{table*}

\begin{figure}
\center
\resizebox{0.45\textwidth}{!}{\includegraphics{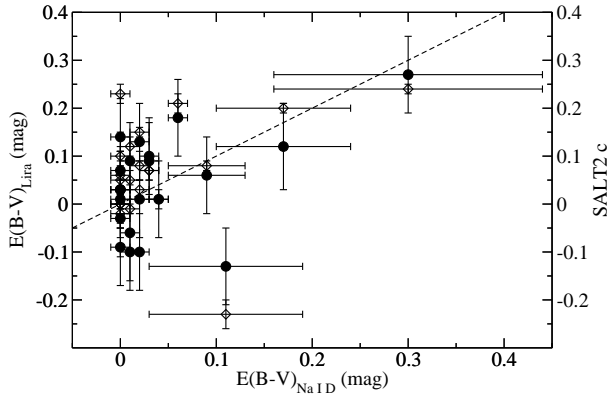}}
\caption{\small $E(B-V)$ as derived from the Lira relation (filled circles) or
the SALT2 $c$ parameter (open diamonds) as fit from SNIFS spectrophotometry,
vs. $E(B-V)$ as derived from the equivalent width of \na\ absorption
\citep[][filled circles]{ppb12}.  The dotted line shows
$E(B-V)_\mathrm{Lira} = E(B-V)_\mathrm{Na~I~D}$.}
\label{fig:ewna-lira}
\end{figure}

Since the Lira relation uses the same late-time data as our mass
reconstruction analysis, it can serve as a separate consistency check on our
data quality.  If a \revised{supernova} has a Lira excess inconsistent with
the extinction implied by \na\ absorption, this could signal a problem with
the late time data (e.g., residual host galaxy contamination).
Figure~\ref{fig:ewna-lira} plots Lira excess against reddening
derived from \ewna\ \revised{and against} SALT2 $c$.
\snf{070902-018} shows up as an outlier with
\mbox{$E(B-V)_\mathrm{Lira} = -0.14 \pm 0.08$~\revised{mag}},
in rough agreement with $c = -0.23 \pm 0.03$, but
\mbox{$E(B-V)_\mathrm{\na} = 0.11^{+0.11}_{-0.06}$~\revised{mag}}.
Since \mbox{$E(B-V)_\mathrm{\na}$} is different from zero at less than 95\%
confidence, \snf{070902-018} could simply have scattered left on the diagram,
or could have \na\ absorption not associated with dust extinction.
For the other SNe, the two reddening estimates
are consistent with each other within the errors, given the substantial
spread of the extinction relations\revised{.  Most of our sample shows
evidence}
for little or no host galaxy extinction.  The \revised{reddening estimates
also track SALT2~$c$} within the uncertainties.


\subsection{Near-Infrared Corrections}
\label{subsec:nircorr}

Since SNIFS observes only wavelengths from 3300--9700~\AA, some fraction of
the bolometric flux at near-infrared wavelengths will be lost.
We correct for this fraction using mean time-dependent corrections derived
from near-infrared $YJHK$ photometry of normal SNe~Ia from
the Carnegie Supernova Project \citep[CSP;][]{csp10,csp11}.

We start with the 67 SNe~Ia published in CSP DR2 \citep{csp11}.  To minimize
the impact of dust extinction, we remove 16 SNe that have SALT2 $c > 0.15$
and are therefore likely to suffer significant host reddening
(including the highly extinguished SN~2006X).
We also remove two superluminous SNe~Ia, SN~2007if \citep{scalzo10,yuan10}
and SN~2009dc \citep{silverman11,taub11}.

For the remaining 49 CSP SNe~Ia, we perform \revised{Gaussian process (GP)}
regression to predict the $YJHK$ magnitudes between rest-frame $B$-band phases
($-14$~d, $+70$~d).  The GP regression fit
\revised{to all NIR observations of these CSP supernovae}
is then used \revised{as a template} to predict the $YJHK$ magnitudes for
the SNfactory sample.  Before fitting, the CSP light curve in each band
$j \in \{Y, J, H, K\}$ is normalized to the $i$-band flux at first maximum,
$i_\mathrm{max}$, so that the quantity predicted by the fit is
$i_\mathrm{max} - m_j$.  To recover the expected NIR magnitudes for
a SNfactory SN, we measure $i_\mathrm{max}$ and apply the measured value
to the GP predictions.
Normalizing the NIR correction relative to $i$-band, which suffers less
extinction than $B$ or the total $UBVRI$ quasi-bolometric flux, results in
a lower systematic error on the NIR correction than if we normalized it
instead to the $B$-band flux or the quasi-bolometric $UBVRI$ flux.
The GP regression fit in each band is shown in Figure~\ref{fig:nir-gp};
further details on the GP training, e.g. the covariance function,
can be found in Appendix \ref{subsec:gp-nir}.

\begin{figure*}
\center
\resizebox{\textwidth}{!}{\includegraphics{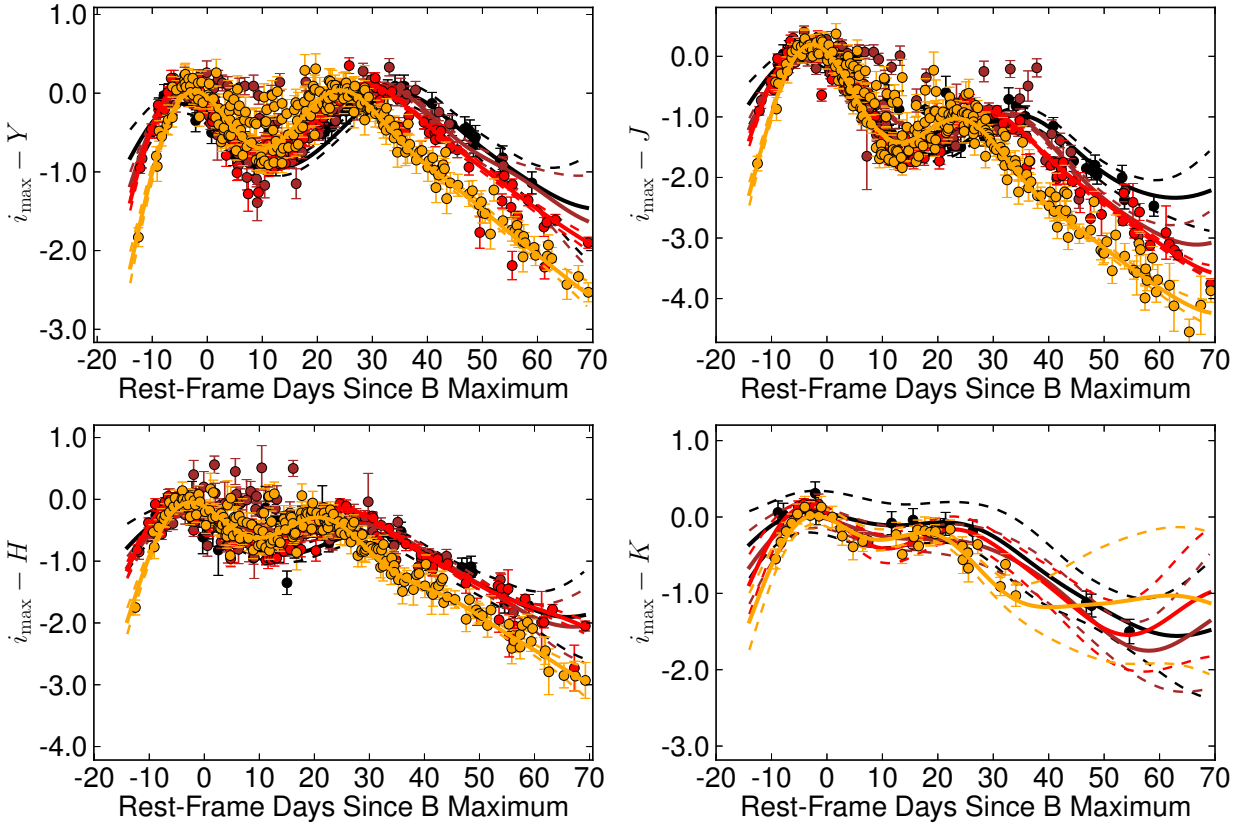}}
\caption{\small Gaussian process regression modeling for $YJHK$ magnitudes
of normal SNe~Ia from the Carnegie Supernova Project \citep{csp10,csp11}.
Bands shown:  $Y$ (upper left), $J$ (upper right), $H$ (lower left),
and $K$ (lower right).  Sections of the GP posterior in ranges of $x_1$
are also shown for each band, along with the CSP data points:
$-2 < x_1 < -1$ (yellow), $-1 < x_1 < 0$ (red),
$0 < x_1 < +1$ (brown), $+1 < x_1 < +2$ (black).}
\label{fig:nir-gp}
\end{figure*}

To generate a bolometric light curve from SNIFS spectrophotometry,
we start with rest-frame, flux-calibrated SNIFS spectra which have been
corrected for Milky Way dust extinction using the \citet{sfd} dust maps
and a \citet{cardelli} reddening law with $R_V = 3.1$.  We first synthesize
the rest-frame $i$-band light curve of the SN and use GP regression
to fit the light curve near maximum light, measuring $i_\mathrm{max}$.
For each SNIFS spectrum, we predict $YJHK$ apparent magnitudes
using the GP regression model with parameters $(x_1, t, i_\mathrm{max})$
as input.  We convert each predicted magnitude $m_j$
to a monochromatic flux density $f_{\lambda_j}$ at the central wavelength of
CSP band $j$:
\begin{equation}
f_{\lambda_j} = 10^{-0.4(m_j-m_{S,j})}
                \frac{\int S(\lambda) T_j(\lambda) \, d\lambda}
                     {\int T_j(\lambda) \, d\lambda}
\end{equation}
where $S(\lambda)$ is the SED of $\alpha$~Lyr \citep{bg04}, with magnitude
$m_{S,j}$ in band $j$ with transmission $T_j(\lambda)$.  We then interpolate
linearly between these flux densities to produce a low-resolution SED,
which extends the SNIFS SED at wavelengths redder than 8800~\AA\ rest-frame.
We integrate the resulting SED from 3300--23900~\AA\ to produce a bolometric
flux at each phase.

\begin{figure}
\center
\resizebox{0.5\textwidth}{!}{\includegraphics{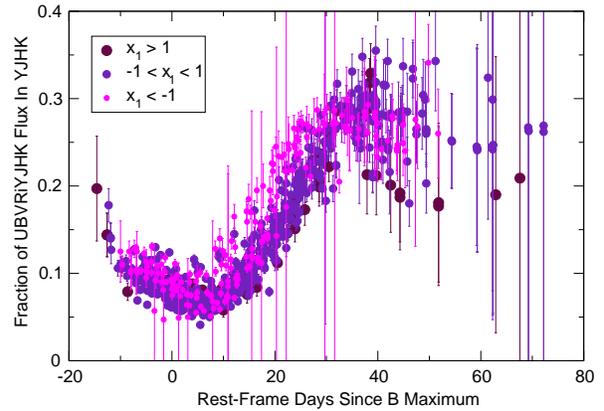}}
\caption{\small NIR correction for unobserved flux in the wavelength range
8800--23900~\AA\ for the SNfactory SNe~Ia, before correcting for host galaxy
extinction.  \revised{
Three ranges in the light curve width parameter $x_1$ are shown here:
fast-declining ($x_1 < -1$, small magenta); average ($-1 < x_1 < 1$, purple);
and slow-declining ($x_1 > 1$, large maroon).}}
\label{fig:fnir}
\end{figure}

The predicted fraction of bolometric flux redward of 8800~\AA\ as a function
of rest-frame $B$-band phase for the SNfactory SNe is presented in
Figure~\ref{fig:fnir}.  While under 10\% near maximum light, the fraction
grows to about 30\% near the NIR second maximum, and then slowly declines.
The fraction is decline-rate dependent, and not negligible at late phases.


\subsection{Final Bolometric Light Curves}
\label{subsec:bololc}

\begin{figure}
\center
\resizebox{0.5\textwidth}{!}{\includegraphics{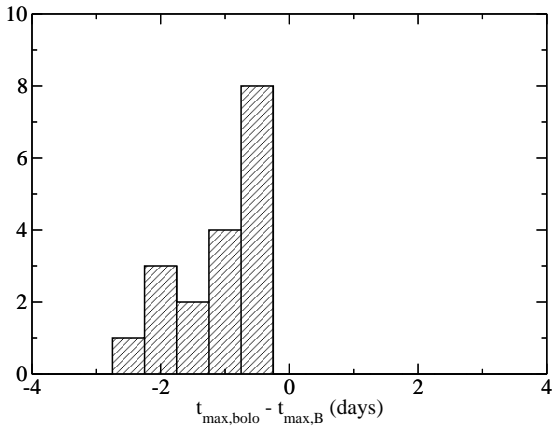}}
\caption{\small Difference between dates of bolometric maximum and $B$-band
maximum for 19 SNe~Ia in our sample.}
\label{fig:bolo-max}
\end{figure}

For each SN in our sample, we generate a series of bolometric light curves
corresponding to different assumptions about host galaxy reddening.
Using a Cardelli extinction law with $R_V = 3.1$ and assumed values of
$E(B-V)_\mathrm{host}$ in 0.01~mag steps from zero to 0.40~mag,
we de-redden the SNIFS spectra before performing the integration and
NIR correction mentioned in \S\ref{subsec:nircorr}.
The ejected mass reconstruction (see \S\ref{sec:modeling}) marginalizes
(integrates) the posterior probability over values of the host galaxy
reddening subject
to a Gaussian prior given by the constraints in Table~\ref{tbl:hostred}.

To ensure that all light curves in our sample have coverage at epochs
appropriate for our modeling, we use a GP regression fit to the bolometric
flux to extract the date of bolometric maximum light and the maximum
bolometric flux.  We use the fitted bolometric maximum flux to constrain
the \nickel\ mass in our reconstruction.

Figure~\ref{fig:bolo-max} shows a histogram of the dates of bolometric maximum
light, relative to the respective dates of $B$-band maximum light from the
SALT2 fit, for the SNe in our sample.  Four of our SNe
(\snf{061020-000}, \snf{070817-003}, \snf{080522-011}, and \snf{080620-000})
have poor constraints on the date of bolometric maximum light from the GP fit,
due to broad-topped light curves or too few early points with full
wavelength coverage; however, their dates of $B$-band maximum light are
well-constrained via SALT2, using information from multiple bands.
For these SNe, we fix the date of bolometric maximum light to equal
$B$-band maximum minus 1 day.
(The mean of the distribution is $-1.1$~days; the median is $-0.9$~days.)

We use independent Cepheid distance estimates to determine the distance
moduli when they are available (SN~2005el, SN~2008ec, SN~2011fe).
For the other SNe, we derive a distance modulus for each SN from its
CMB-centric host galaxy redshift assuming a \lcdm\ cosmology
($\Omega_\mathrm{M} = 0.28$, $\Omega_\Lambda = 0.72$,
$H_0 = 72$~\kms~Mpc$^{-1}$).  The resulting absolute bolometric light curves
are the input to our mass reconstruction in \S\ref{sec:modeling}.


\section{Modeling}
\label{sec:modeling}

For reconstruction of \nickel\ masses and ejected masses of the SNfactory
SNe~Ia, we use a new implementation of the Markov chain Monte Carlo code
featured in \citet{scalzo12}, to which we refer the interested reader for
a more detailed discussion of the physics involved.  We summarize the overall
method briefly here in \S\ref{subsec:MassSolver}, and describe our fiducial
set of priors in \S\ref{subsec:assumptions}.  We test the code on a suite of
contemporary SN~Ia explosion models in \S\ref{subsec:mpa-challenge} and
discuss features of the dependence of the bolometric light curve on the
physical parameters of the system in \S\ref{subsec:bololc-comp},
before discussing application of the method to SNfactory observations
in subsequent subsections.


\subsection{The Reconstruction Method}
\label{subsec:MassSolver}

Our reconstruction code calculates the late-time bolometric light
curve in the optically thin limit of Compton scattering of gamma rays from
\cobalt\ decay.  The code fits two parameters, a \nickel\ mass \MNi\ and
a fiducial time $t_0$ at which the optical depth to Compton scattering
equals unity, using Arnett's rule \citep{arnett82} and the analytic treatment
of \citet{jeffery99}.  The \nickel\ mass is calculated via
\begin{equation}
\MNi = \frac{L_\mathrm{bol,max}}{\alpha \dot{S}(t_{R,\mathrm{bol}})},
\end{equation}
where $L_\mathrm{bol,max}$ is the maximum bolometric luminosity,
$t_{R,\mathrm{bol}}$ is the rise time to bolometric maximum,
$\dot{S}(t)$ is the instantaneous rate of radioactive energy release
from the \nickel\ decay chain at time $t$ since explosion, and
$\alpha$ is a model-dependent dimensionless number of order unity related
to the diffusion time of radiation through the ejecta at early times.
The transparency time $t_0$ at late times is calculated from
\begin{equation}
L_\mathrm{bol}(t) = \left[ 1-e^{-(t/t_0)^{-2}} \right]
   \dot{S}_\gamma(t) + \dot{S}_{e^+}(t)
\end{equation}
where we have now split $S(t) = S_\gamma(t) + S_{e^+}(t)$ into the radioactive
energy release from gamma rays, some of which will escape the ejecta,
and from positrons, which we treat as fully trapped at this stage of
evolution of the expanding SN remnant ($t < 120$~days after explosion).
Note that $\alpha$ does not appear in the late-time expression; it includes
reprocessing of radiation at gamma-ray and at optical wavelengths, but at
late times trapping of optical radiation is much reduced and
changes in gamma-ray transparency are encoded in $t_0$.

To first order, then, \MNi\ controls the overall level of radioactivity and
determines the overall flux scale of the light curve, while $t_0$ controls
the rate at which the radiation escapes from the ejecta and hence the shape
of the light curve.
We then map these two numbers, \MNi\ and $t_0$, to a total ejected mass
\MWD\ using a Monte Carlo Markov chain (MCMC).  The configuration of the
model system is described by a total mass \MWD, a velocity scale \vKE,
a central density $\rho_c$, a composition
$(f_\mathrm{Fe}, f_\mathrm{\nickel}, f_\mathrm{Si}, f_\mathrm{CO})$,
and nuisance parameters
$(\alpha, \aNi, t_{R,\mathrm{bol}}, E(B-V)_\mathrm{host})$ subject to
the following prior constraints:
\begin{enumerate}
\item the density structure is a spherically symmetric function of velocity
      $\rho(v/\vKE)$;
\item the value of \vKE\ is set, for a given composition, via conservation
      of energy, by constraining the kinetic energy
      $E_K = \frac{1}{2}\MWD \vKE^2$ to be the difference between the
      nuclear energy $E_N$ \citep{mi09} released in the explosion
      and the binding energy
      $E_G$ of a white dwarf of mass \MWD\ and central density $\rho_c$;
\item we use the binding energy formula of \citet{yl05}, which has been
      used elsewhere to account for the angular momentum of rotating
      super-Chandrasekhar-mass white dwarfs
      \citep{howell06,jbb06,mi09,scalzo10,scalzo12}, and which reduces to the
      usual non-rotating formula for sub-Chandrasekhar-mass white dwarfs;
\item the ratio $\eta = \MNi/(\MNi + M_\mathrm{Fe})$ of \nickel\ to overall
      iron-peak element yield is a function of $\rho_c$
      \citep{krueger10,krueger12,seitenzahl11}, with higher central
      densities resulting in more neutronization and a higher fraction
      of stable iron-peak elements;
\item mixing of \nickel\ through the ejecta is set by a mixing parameter
      $a_\nickel$ \citep{kasen06} which describes the scale over which
      mixing takes place in enclosed mass coordinates
      $m(v) = \MWD^{-1} \int_0^v 4\pi v^2 \rho(v) \, dv$.
\end{enumerate}
The ejected mass itself satisfies
\begin{equation}
\MWD = \frac{4\pi}{\kappa_\gamma Q} (\vKE t_0)^2,
\label{eqn:MWD}
\end{equation}
where $\kappa_\gamma$ is the effective opacity of the ejecta to Compton
scattering, and $Q$ is a form factor describing the
\nickel-weighted Compton scattering optical depth for the given density
profile and ejecta composition, similar to $q$ in \citep{jeffery99}.
For a density profile with an exponential dependence on velocity,
the case treated explicitly in \citet{jeffery99}
\revised{and in \citet{stritz06}}, $Q = 6q$.
We populate a look-up table for $Q$ as a function of the ejecta composition
by numerically evaluating the necessary integrals using the VEGAS algorithm
\cite{vegas}, as in \citet{scalzo12}.

We use the parallel-tempered \revised{MCMC} sampler
\texttt{emcee} \citep{emcee}, which simultaneously runs several ensembles
of ``walkers'' with different step sizes (``temperatures'') and shares
information between them.  This method is appropriate for likelihood surfaces
with multiple maxima, which may be the case for our problem ---
for example, a fast-declining light curve could in principle be described
by a low-\MWD\ solution with a \nickel\ distribution strongly concentrated
at the centre, or by a high-\MWD\ solution in which the \nickel\ lies closer
to the surface.  We verify that convergence has been reached by comparing
runs of different lengths.
In general we find a ``burn-in'' period of 1500 iterations, which are then
discarded, suffices to remove dependence on the initial conditions.
Our results are then obtained by sampling for an additional $1500 \times k$
iterations, recording every $k$th iteration
where $k$ is the autocorrelation time in iterations of the chain.
Our final probability distributions contain about $3 \times 10^5$ samples
over all parameter configurations for each SN.


\subsection{Fiducial Priors for Normal SNe~Ia}
\label{subsec:assumptions}

Although the capabilities of the modeling code as used in this paper are
the same as in \citet{scalzo12}, we use a set of priors more appropriate
for normal SNe~Ia, rather than 1991T-like or super-Chandrasekhar SNe~Ia.
We describe these assumptions here.

Consistent with our previous work \citep{scalzo10,scalzo12}, we adopt
the prior $\kappa_\gamma = 0.025$~cm$^2$~g$^{-1}$ \citep{swartz95,jeffery99},
as appropriate for the case of Compton-thin ejecta.  This number allows us
to accurately convert from a measured column density for Compton scattering
to the mass of ejecta.  Most of our other priors below are targeted at making
a reasonable guess about the \emph{distribution} of \nickel\ in the ejecta,
which will affect our results through the form factor $Q$.

While $\alpha = 1.2$ is a common choice when deriving \MNi\ for SNe~Ia
\citep{nugent95,jbb06,howell06,howell09}, there is some uncertainty in its
true value.  The self-consistent, albeit simple, model of \citet{arnett82}
accounts for radiation trapping and has $\alpha$ very close to 1.0.
The models of \citet{hk96} cover the range 0.8--1.6 with a mean of 1.0
and a standard deviation of 0.2.  Some other analyses also fix $\alpha = 1.0$
explicitly \citep[e.g.][]{stritz06,mazzali07}.  For compatibility with a
broad range of explosion scenarios, we choose $\alpha = 1.2 \pm 0.2$ for our
fiducial analysis.  However, we also run \revised{reconstructions} with fixed
$\alpha = 1.0$, for comparison with some of the previous literature, and
to estimate how much of our final error budget results from uncertainty
in the true value of $\alpha$ as derived from full simulations.

The rise time and $B$-band decline rate of normal SNe~Ia are strongly
correlated \citep{ganesh11}, and since the date of bolometric maximum is
strongly tied to that of $B$-band maximum, we use this information to
estimate the bolometric rise time
\begin{equation}
t_{R,\mathrm{bol}} = t_{R,B} + (t_\mathrm{max,bol} - t_{\mathrm{max},B})
\end{equation}
by extracting the dates $t_\mathrm{max,bol}$ and $t_{\mathrm{max},B}$
of maximum light of the bolometric and $B$-band light curves from the
respective GP fits to those light curves.  We estimate
the $B$-band rise time via the relation
\begin{equation}
t_{R,B} = 17.5 - 5(\Delta m_{15,B} - 1.1)~\mathrm{days}
\end{equation}
which covers the $t_{R,B}$ vs. $\Delta m_{15,B}$ locus of \citet{ganesh11};
we assign a relatively conservative error of $\pm 2$~days to this estimate.
We find that bolometric maximum light precedes $B$-band maximum light by
about 1 day on \revised{an} average for the SNe in our sample, so our prior on
$t_{R,B}$ translates to $t_{R,\mathrm{bol}} = 16.5 \pm 2$~days in practice
for a typical SN~Ia with $\Delta m_{15,B} = 1.1$ \mbox{(SALT2 $x_1 = 0$)}.
For those SNe for which $B$-band maximum was fixed and not directly
observed, we increase the uncertainty in the rise time to $\pm 3$~days
(the spread from Figure~\ref{fig:bolo-max}).

The central density $\rho_c$ of the progenitor at the time of explosion
influences our results through the binding energy
(affecting the kinetic energy of the ejecta) and through neutronization
(affecting the mass fraction of stable iron-peak elements).
\citet{seitenzahl09} investigate the criteria for the formation of a
detonation, and find that they may occur at densities as low as
$3 \times 10^6$~g~cm$^{-3}$, while the lowest-mass white dwarf considered
in \citet{fink10} had a central density of $1.4 \times 10^7$~g~cm$^{-3}$.
At densities of $10^{10}$~g~cm$^{-3}$ or higher, accretion-induced collapse
(AIC) to a neutron star is more likely than a SN~Ia explosion \citep{nk91}.
However, recent studies investigating the extent of neutronization in
delayed detonation simulations of SN~Ia explosions
\citep{krueger10,krueger12,seitenzahl11}, which inform our neutronization
prior (see below), do not consider $\rho_c > 5 \times 10^9$~g~cm$^{-3}$.
We therefore require $7.0 < \log_{10}~\rho_c < 9.7$, while acknowledging that
solutions with central densities outside this range could in principle exist
and produce normal SNe~Ia.

Since neutronization in the explosion may affect the distribution of \nickel\
in the ejecta and hence the value of $Q$, it is important for our purposes
to account for it somehow.  \citet{krueger10} and \citet{krueger12} use suites
of 2-D simulations to explicitly constrain the dependence of \MNi\ and \MFe\
on $\rho_c$.  \citet{seitenzahl11} use a smaller suite of 3-D simulations to
address the same question, with slightly larger scatter.  While they disagree
on how the overall iron-peak element yield varies with $\rho_c$, the two sets
of models show similar mean behavior of $\eta(\rho_c)$ within the scatter.
We therefore adopt the Gaussian prior
\begin{equation}
\eta = 0.95 - 0.05\,\rho_{c,9} \pm 0.03\,\max(1, \rho_{c,9}),
\end{equation}
with $\rho_{c,9} = \rho_c / 10^9$~g~cm$^{-3}$, which should be consistent with
both sets of simulations; as specified above, we rely on the luminosity of
each SN to constrain the actual value of \MNi.  This is slightly different
than the prior used in \citet{scalzo12}\revised{,} which was
informed only by the results of \citet{krueger10}.

In \citet{scalzo12}, we allowed our composition structure to have central
concentrations of stable iron-peak elements, or central deficits of \nickel,
for explosions of progenitors with high central density, as expected in some
1-D delayed detonation models \citep{kmh93,hk96,blondin13}.
Recent multi-dimensional simulations of delayed detonations
\citep{krueger12,seitenzahl13},
on the other hand, find no evidence for such central \nickel\ deficits:
during the deflagration phase, plumes of hot iron-peak ash rise through the
ejecta rather than remaining centrally concentrated, a behavior which cannot
take place in 1-D hydrodynamic models.  On average, the resulting composition
structure is consistent with an approximately constant ratio of \nickel\
to stable iron-peak elements throughout the ejecta.  Under this (reasonable)
assumption, we find that the dependence of $Q$ on the stable iron-peak content
of the ejecta is much reduced, leading to tighter constraints on the ejected
mass.  We therefore choose a case with no central \nickel\ hole as our
fiducial analysis.  For completeness, however, we shall also explore the
influence of a \nickel\ hole.  Some 3-D models, such as the violent
double-degenerate mergers of \citet{pakmor12}, show \nickel\ holes due simply
to the dynamics of the merger and not due to neutronization.

We choose $\aNi = 0.2$, typical of the ``moderate mixing'' case shown in
\citet{kasen06}.  We expect that this value will reproduce the near-infrared
light curves of the typical normal SN~Ia, with two distinct maxima, better
than the ``enhanced mixing'' case $\aNi = 0.5$, which results in a strongly
suppressed second maximum typical of overluminous supernovae such as the
super-Chandrasekhar-mass candidates presented in \citet{scalzo12}.
While there may be some variation in the true value of \aNi\ throughout the
population, we use $\aNi = 0.2$ as a representative value.  In future
investigations the morphology of the near-infrared light curve could in
principle be used to constrain \aNi.

While it may be tempting to try to constrain \vKE\ by using \ion{Si}{2}
information near maximum light, we choose not to do so here.
In \citet{scalzo10} and \citet{scalzo12}, we used \ion{Si}{2} absorption
minimum velocities near maximum light to constrain the mass of the
reverse-shock shell in a \revised{``}tamped-detonation\revised{''} scenario
\revised{\citep{kmh93,hk96}, in which the supernova ejecta interact with
a dense carbon/oxygen envelope characteristic of double-degenerate mergers.}
However, the presence of the shell immediately implied that the photospheric
velocity matched the velocity of the disturbed outer ejecta, and had no
bearing on the kinetic energy scale of the bulk ejecta most relevant for
the gamma-ray transparency measurement of the ejected mass.  Even for SNe
with smoother density structures, a variety of velocities and velocity
gradients may be possible \citep[e.g.][]{blondin11,blondin13}.
While comparison to detailed radiation transfer models could provide
constraints on \vKE\ from photospheric velocities, it is beyond the capacity
of our current semi-analytic treatment.  However, our model self-consistently
predicts \vKE\ as a function of mass, central density, and composition.
We typically obtain $\vKE \sim 10500$~\kms, a plausible value for SNe~Ia.

We limit the mass of unburned carbon and oxygen $M_\mathrm{CO}/\MWD < 0.05$,
since carbon is rarely seen in SNe~Ia except in spectra taken a week or
more before maximum light \citep{rct07,rct11,folatelli11}.
This results in a constraint on \vKE\ and rules out models with large
amounts of unburned carbon and oxygen but no intermediate-mass elements.
While we use this constraint in our fiducial analysis, we will also present
results without this constraint later.

\begin{figure}
\center
\resizebox{0.5\textwidth}{!}{\includegraphics{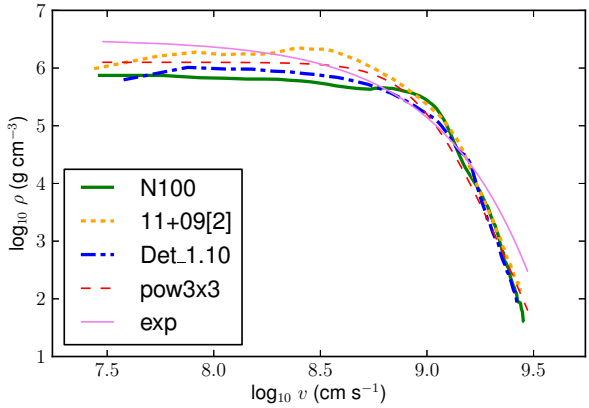}}
\caption{\small The ``exp'' and ``pow3x3'' density profiles, along with the
angle-averaged density profiles $\rho(v)$ for three 3-D explosion models:
N100 \citep{seitenzahl13}, 11+09 \citep{pakmor12},
and Det\_1.10 \citep{ruiter13}.}
\label{fig:density-profiles}
\end{figure}

Finally, the choice of density profile also affects the inferred mass
through $Q$, and this choice can be informed only by hydrodynamic simulations
of SN explosions.  We consider two possible density profiles.
An exponential density profile $\rho(v) \propto \exp(-\sqrt{12}v/\vKE)$
(``exp'') is a good description of many 1-D explosion models
\citep{w7,kmh93,hk96,blondin13} and a mathematically convenient assumption
in previous SN~Ia work \citep{jeffery99,stritz06,jbb06,kasen06}.
For consistency with this prior work we use an exponential density profile
in our fiducial analysis.  However, our framework is flexible and allows for
arbitrary density profiles, so here we also consider
$\rho(v) \propto [1 + (v/\vKE)^{3}]^{-3}$ (``pow3x3''), which reduces to a
power law $v^{-9}$ at large velocities.  The ``pow3x3'' profile was chosen
specifically to provide a structure representative of the 3-D explosion
models discussed in \S\ref{subsec:mpa-challenge} below.  A visual comparison
of the density profiles of representative explosion models with our density
profiles of choice is shown in Figure~\ref{fig:density-profiles}.
We could also consider highly disturbed density profiles appropriate
to tamped detonations or pulsating delayed detonations \citep{kmh93,hk96},
as in our previous work on candidate super-Chandrasekhar-mass SNe~Ia
\citep{scalzo10,scalzo12}.  However, the late-time bolometric light curve is
sensitive mainly to the overall column density presented to outbound \cobalt\
gamma rays (i.e., on $Q$)\revised{.  A} density enhancement due to a
shock in the outer layers will not influence $Q$ as long as it does not
extend into the \nickel-rich inner ejecta.

The results of the mass reconstruction for our fiducial analysis are shown
in Table~\ref{tbl:massrecon}.  Since the probability distributions of the
tabulated quantities are significantly non-Gaussian, the (asymmetric) error
bars we quote bound the 68\% confidence region.  We also tabulate the
probability $P(>M_\mathrm{Ch})$ that the SN's mass exceeds 1.4~\Msol,
very high or low values of which indicate significant deviation from a
Chandrasekhar-mass explosion.

\begin{table*}
\caption{Mass reconstruction of SNfactory bolometric light curves}
\newcommand{\aatag}{\ensuremath{^\dagger}}
\begin{tabular}{lrrrrrr}
\hline 
   SN~Name &
   ${\MWD/\Msol}\nb{a}$ &
   ${\MNi/\Msol}\nb{b}$ &
   ${t_0}\nb{c}$~(days) &
   ${P_\mathrm{SCh}}\nb{d}$ &
   ${P_\mathrm{fit}}\nb{e}$ \\
\hline 
\multicolumn{6}{c}{SNfactory-Discovered Supernovae} \\
\hline 
SNF~20060907-000 & $1.01^{+0.09}_{-0.07}$ & $0.56 \pm 0.12$ & $33.8 \pm 4.1$ & $0.001$ & $0.797$ \\[0.2ex]
SNF~20061020-000 & $0.99^{+0.11}_{-0.09}$ & $0.34 \pm 0.09$ & $37.8 \pm 4.4$ & $0.002$ & $0.355$ \\[0.2ex]
SNF~20070506-006\aatag & $1.53^{+0.17}_{-0.11}$ & $0.71 \pm 0.14$ & $47.4 \pm 5.8$ & $0.885$ & $0.788$ \\[0.2ex]
SNF~20070701-005 & $1.31^{+0.11}_{-0.10}$ & $0.83 \pm 0.17$ & $38.3 \pm 4.1$ & $0.224$ & $0.438$ \\[0.2ex]
SNF~20070810-004 & $1.35^{+0.15}_{-0.17}$ & $0.40 \pm 0.08$ & $47.3 \pm 6.3$ & $0.392$ & $0.730$ \\[0.2ex]
SNF~20070817-003 & $1.04^{+0.12}_{-0.10}$ & $0.33 \pm 0.09$ & $39.6 \pm 4.8$ & $0.011$ & $0.717$ \\[0.2ex]
SNF~20070902-018 & $1.18^{+0.15}_{-0.13}$ & $0.36 \pm 0.08$ & $43.1 \pm 5.2$ & $0.081$ & $0.364$ \\[0.2ex]
SNF~20080522-011 & $1.40^{+0.12}_{-0.12}$ & $0.61 \pm 0.15$ & $45.0 \pm 5.7$ & $0.518$ & $0.355$ \\[0.2ex]
SNF~20080620-000 & $1.14^{+0.16}_{-0.12}$ & $0.32 \pm 0.07$ & $42.7 \pm 5.2$ & $0.070$ & $0.775$ \\[0.2ex]
SNF~20080717-000 & $1.46^{+0.12}_{-0.09}$ & $0.80 \pm 0.20$ & $43.3 \pm 4.9$ & $0.735$ & $0.204$ \\[0.2ex]
SNF~20080803-000 & $1.34^{+0.13}_{-0.13}$ & $0.61 \pm 0.15$ & $42.5 \pm 5.4$ & $0.333$ & $0.711$ \\[0.2ex]
SNF~20080913-031 & $1.10^{+0.12}_{-0.10}$ & $0.43 \pm 0.09$ & $39.2 \pm 4.7$ & $0.015$ & $0.782$ \\[0.2ex]
SNF~20080918-004 & $0.92^{+0.08}_{-0.06}$ & $0.30 \pm 0.05$ & $36.4 \pm 3.0$ & $0.000$ & $0.733$ \\[0.2ex]
\hline 
\multicolumn{6}{c}{Externally Discovered Supernovae Observed by SNfactory} \\
\hline 
SN~2005el        & $0.90^{+0.06}_{-0.05}$ & $0.52 \pm 0.12$ & $31.4 \pm 3.0$ & $0.000$ & $0.570$ \\[0.2ex]
SN~2007cq        & $1.17^{+0.12}_{-0.10}$ & $0.53 \pm 0.12$ & $39.4 \pm 4.9$ & $0.046$ & $0.738$ \\[0.2ex]
SN~2008ec        & $1.02^{+0.10}_{-0.09}$ & $0.34 \pm 0.08$ & $38.5 \pm 4.0$ & $0.002$ & $0.506$ \\[0.2ex]
SN~2011fe        & $1.19^{+0.12}_{-0.11}$ & $0.42 \pm 0.08$ & $42.4 \pm 4.5$ & $0.057$ & $0.585$ \\[0.2ex]
PTF09dlc         & $1.24^{+0.14}_{-0.11}$ & $0.48 \pm 0.10$ & $42.4 \pm 5.3$ & $0.129$ & $0.772$ \\[0.2ex]
PTF09dnl         & $1.33^{+0.13}_{-0.13}$ & $0.48 \pm 0.10$ & $45.2 \pm 5.3$ & $0.324$ & $0.509$ \\[0.2ex]
\hline 
\end{tabular}
\medskip \\
\flushleft
Quantities with error bars are marginalized over all independent parameters.  Uncertainties represent
the 68\% CL intervals for the projections of the multi-dimensional PDF of the fiducial analysis
onto the derived quantities.
Fiducial priors:  $\rho(v) \sim \exp(-\sqrt{12}v/\vKE)$, $\alpha = 1.2 \pm 0.2$, no \nickel\ hole. \\
\nb{a}~{Total ejected mass.} \\
\nb{b}~{\nickel\ mass synthesized in the explosion.}\\
\nb{c}~{Time since explosion, in days, at which $\tau = 1$
        for Compton scattering of \cobalt\ gamma rays in the ejecta.}\\
\nb{d}~{Fraction of the integrated probability density lying above $\MWD = 1.4$~\Msol.}\\
\nb{e}~{Probability of attaining the given value of $\chi^2_\nu$ or higher
      if the model is a good fit to the data, incorporating all priors.}\\
\aatag~Typed by SNID as 1999aa-like from multiple pre-maximum spectra.

\label{tbl:massrecon}
\end{table*}


\subsection{Reconstruction of Simulated Light Curves}
\label{subsec:mpa-challenge}

As a test of the code, we have run our reconstruction \revised{code} on
a set of simulated bolometric light curves \revised{of} numerical explosion
models generated with the Monte Carlo
radiation transfer code ARTIS \citep{ks09}.  The models span a range of
masses from 1.06~\Msol\ to 1.95~\Msol\ and different explosion
mechanisms, and provide synthetic observables from well before bolometric
maximum to about 75 days after bolometric maximum.  We assign an error of
0.03~mag to each point, although the actual light curves have much lower
statistical noise; this represents approximately what our
method could achieve in the limit of very high signal-to-noise.
The models were reconstructed in a blind analysis, using the same input
assumptions as our fiducial analysis (``Run~A'' in \S\ref{subsec:xchckes})
on the SNfactory sample, with the model identities and true ejected masses
and \nickel\ masses unknown until the reconstruction had been performed.

The results of the reconstruction are shown in Table~\ref{tbl:massrecon-mpa},
along with the unblinded model identities and references.
We remind the reader that our Monte Carlo sampler does not search for a
single set of best-fitting parameters for a given light curve, but samples
the entire probability distribution of allowed parameter values.
The columns in the table represent projections of this probability
distribution onto the variables of interest, marginalizing
(i.e. \revised{integrating}) over all other variables.
Since the probability distributions for the reconstructed quantities are in
general asymmetric with non-Gaussian tails, we quote the median value as
the central value estimate with the 68\% confidence intervals expressed as
asymmetric error bars, and also show the total integrated probability of
the reconstructed parameters above $M = M_\mathrm{Ch} = 1.4~\Msol$.

\begin{table*}
\caption{Mass reconstruction of simulated bolometric light curves}
\begin{tabular}{lccrrrrrr}
\hline
   & \multicolumn{2}{c}{True Parameters} & \multicolumn{3}{c}{Reconstructed Parameters} & & \\
   SN~Name &
   ${\MWD/\Msol}^\mathrm{a}$ &
   ${\MNi/\Msol}^\mathrm{b}$ &
   ${\MWD/\Msol}^\mathrm{a}$ &
   ${\MNi/\Msol}^\mathrm{b}$ &
   ${t_0}^\mathrm{c}$~(days) &
   ${P_\mathrm{SCh}}^\mathrm{d}$ &
   ${P_\mathrm{fit}}^\mathrm{e}$ \\
\hline
\multicolumn{8}{c}{Using Full Late-Time Light Curve} \\
\hline
Model 3\nb{f}   & 1.07 & 0.60 & $1.01^{+0.09}_{-0.08}$ & $0.34 \pm 0.05$ & $38.3 \pm 3.5$ & $0.002$ & $0.577$ \\[0.2ex]
Det\_1.10\nb{g} & 1.10 & 0.62 & $1.22^{+0.12}_{-0.11}$ & $0.38 \pm 0.06$ & $44.1 \pm 4.3$ & $0.089$ & $0.579$ \\[0.2ex]
N5\nb{h}        & 1.40 & 0.97 & $1.35^{+0.11}_{-0.11}$ & $0.60 \pm 0.10$ & $43.5 \pm 4.6$ & $0.331$ & $0.712$ \\[0.2ex]
N100\nb{h}      & 1.40 & 0.60 & $1.27^{+0.14}_{-0.12}$ & $0.40 \pm 0.07$ & $45.0 \pm 5.0$ & $0.197$ & $0.917$ \\[0.2ex]
N1600\nb{h}     & 1.40 & 0.32 & $1.46^{+0.18}_{-0.11}$ & $0.21 \pm 0.02$ & $55.9 \pm 4.3$ & $0.713$ & $0.689$ \\[0.2ex]
11+09[1]\nb{i}  & 1.95 & 0.62 & $1.87^{+0.38}_{-0.18}$ & $0.42 \pm 0.05$ & $64.7 \pm 5.8$ & $1.000$ & $0.890$ \\[0.2ex]
11+09[2]\nb{i}  & 1.95 & 0.62 & $1.66^{+0.15}_{-0.10}$ & $0.91 \pm 0.14$ & $47.6 \pm 4.7$ & $1.000$ & $0.662$ \\[0.2ex]
11+09[3]\nb{i}  & 1.95 & 0.62 & $1.59^{+0.22}_{-0.13}$ & $0.37 \pm 0.04$ & $57.1 \pm 4.6$ & $0.957$ & $0.792$ \\[0.2ex]
\hline
\multicolumn{8}{c}{Using Only Data at +40 Days} \\
\hline
Model 3\nb{f}   & 1.07 & 0.60 & $1.08^{+0.11}_{-0.10}$ & $0.34 \pm 0.05$ & $39.2 \pm 4.1$ & $0.005$ & $0.669$ \\[0.2ex]
Det\_1.10\nb{g} & 1.10 & 0.62 & $1.17^{+0.12}_{-0.11}$ & $0.41 \pm 0.07$ & $40.8 \pm 4.6$ & $0.047$ & $0.810$ \\[0.2ex]
N5\nb{h}        & 1.40 & 0.97 & $1.33^{+0.11}_{-0.11}$ & $0.63 \pm 0.11$ & $40.8 \pm 4.7$ & $0.300$ & $0.786$ \\[0.2ex]
N100\nb{h}      & 1.40 & 0.60 & $1.28^{+0.14}_{-0.13}$ & $0.41 \pm 0.07$ & $43.7 \pm 5.2$ & $0.218$ & $0.764$ \\[0.2ex]
N1600\nb{h}     & 1.40 & 0.32 & $1.41^{+0.16}_{-0.17}$ & $0.23 \pm 0.03$ & $51.9 \pm 5.6$ & $0.521$ & $0.679$ \\[0.2ex]
11+09[1]\nb{i}  & 1.95 & 0.62 & $2.00^{+0.57}_{-0.30}$ & $0.42 \pm 0.07$ & $65.9 \pm 10.3$ & $0.999$ & $0.752$ \\[0.2ex]
11+09[2]\nb{i}  & 1.95 & 0.62 & $1.66^{+0.15}_{-0.10}$ & $0.97 \pm 0.15$ & $44.9 \pm 4.9$ & $1.000$ & $0.637$ \\[0.2ex]
11+09[3]\nb{i}  & 1.95 & 0.62 & $1.76^{+0.46}_{-0.24}$ & $0.34 \pm 0.05$ & $60.8 \pm 8.9$ & $0.972$ & $0.740$ \\[0.2ex]
\hline
\end{tabular}
\medskip \\
\flushleft
Quantities with error bars are marginalized over all independent parameters.  Uncertainties represent
the 68\% CL intervals for projections of the multi-dimensional PDF of the fiducial analysis
(the original blind test of the reconstruction method) onto the derived quantities.
Fiducial priors:  $\rho(v) \sim \exp(-\sqrt{12}v/\vKE)$, $\alpha = 1.2 \pm 0.2$, no \nickel\ hole. \\
\nb{a}~{Total ejected mass.}\\
\nb{b}~{\nickel\ mass synthesized in the explosion.}\\
\nb{c}~{Time since explosion, in days, at which $\tau = 1$
        for Compton scattering of \cobalt\ gamma rays in the ejecta.}\\
\nb{d}~{Fraction of the integrated probability density lying above $\MWD = 1.4$~\Msol.}\\
\nb{e}~{Probability of the model is a good fit to the data, incorporating all priors.}\\
\nb{f}~{Reference:  \cite{kromer10}.}\\ 
\nb{g}~{Reference:  \cite{ruiter13}.}\\
\nb{h}~{Reference:  \cite{seitenzahl13}.}\\ 
\nb{i}~{Reference:  \cite{pakmor12}.  Reconstructions from three different views are shown:
        1 = angle-averaged light curve, 2 = brightest line of sight, 3 = faintest line of sight.}\\

\label{tbl:massrecon-mpa}
\end{table*}

The reconstructed masses agree surprisingly well with the model masses,
given that the input assumptions were not tuned to match the explosion models.
In general the reduced chi-squares are modest, showing that
the \citet{jeffery99} functional form can provide a good description of
the simulated light curves within the time range in which it applies.
The true ejected mass lies within the formal 68\% confidence interval on \MWD\
for five of the eight cases, and within the 95\% confidence interval for all
eight cases.
Just as importantly for our purposes, except for the sub-Chandrasekhar model
Det\_1.10, the code correctly distinguishes the non-Chandrasekhar-mass models
at high significance ($> 95\%$~CL) from the Chandrasekhar-mass models.

Three of the light curves represent different lines of sight for the same
violent merger model 11+09, with $\MWD = 1.95~\Msol$, $\MNi = 0.62~\Msol$
\citep{pakmor12}:  the angle-averaged light curve and the brightest and
faintest viewing angles.  Our method gives a very accurate result for the
angle-averaged light curve, but slightly underestimates the ejected mass in
both asymmetric views.  However, in each case it still correctly identifies
the event as super-Chandrasekhar at high ($> 95\%$~CL) significance.
The angle-averaged \nickel\ fraction has a hole in the centre
\citep[see figure 2 of][]{pakmor12}, though it originates from an interaction
with the secondary star rather than neutronization.  When this is accounted
for in our \revised{priors}, the reconstructed masses of versions 1, 2, and 3
become $2.31^{+0.26}_{-0.37}~\Msol$, $1.83^{+0.37}_{-0.23}~\Msol$,
and $1.94^{+0.33}_{-0.30}~\Msol$, respectively, \revised{with the true value}
within the 68\% CL interval for \revised{each} reconstruction.

The derived \nickel\ masses are less secure.  They are quite wrong for the
asymmetric views of 11+09, as one might expect since Arnett's rule assumes
spherical ejecta.  This suggests that some, though not necessarily all,
events which appear to have too much \nickel\ for their reconstructed mass
may in fact be bright views of an asymmetric explosion.  In such a scenario
we would expect more variation in the derived \MNi/\MWD\ ratio for low-\nickel\
events.  For models with less pronounced asymmetries, such as the N5, N100
and N1600 delayed detonations, the reconstructed value of \MNi\ is in general
about 50\% lower than the true value.  This is due to a combination of
factors:  the actual value of $\alpha$ is closer to 1.0~in the simulations
than the central value of 1.2 we assume for our prior, and some of the models
(for example, N100) have more high-velocity \nickel\ than we assume,
affecting the interpretation of the late-time light curves.

Since the reconstructed mass distributions are non-Gaussian, the pull
distribution $(\MWD-M_\mathrm{WD,true})/\sigma_{M_\mathrm{WD}}$ will not have
its usual interpretation, but may still be useful as an indication of how far
wrong our reconstructions are\revised{,} and in which direction.  Using the
appropriate one-sided 68\% uncertainty for each object, we find that the pull
distribution has mean $-0.52$ and standard deviation 0.95; an unbiased sample
drawn from a Gaussian should have mean within $[-0.35, 0.35]$ (1$\sigma$) and
width near 1.  Thus, within this small but fairly diverse selection of
explosion models, our baseline assumptions seem to incur only a small bias,
if any.  The uncertainties scale with mass, with sub-Chandrasekhar-mass
reconstructions being the most secure in absolute terms.

Table~\ref{tbl:massrecon-mpa} also includes results where we use only the
first light curve point more than 40 days after bolometric maximum, since many
of our SNe will have only this point at late times.  This makes
the minimum value of $\chi^2/\nu$ meaningless as a hypothesis testing measure,
since the fit will not be overconstrained, but the Monte Carlo sampler will
still be able to use the likelihood to reject models which do not fit
the data.  The results are largely unchanged; the pull distribution is not
dramatically different (mean $-0.35$, standard deviation $0.75$), and the true
ejected masses still lie within the 95\% CL interval for all eight models.
The code also still accurately distinguishes between sub-Chandrasekhar,
Chandrasekhar-mass and super-Chandrasekhar explosions.  While the
reconstruction may therefore be slightly less accurate and/or precise for
SNe~Ia with fewer or less accurate late-time photometry points, the broad
trends of the mass distribution are still preserved.

In summary, while the code does not perform perfectly on every input model,
it does at least seem to provide reasonable estimates of the uncertainties:
62.5\% of the models lie within the 68\% confidence region.  The results give
us some confidence that the method is relatively robust to systematics, and
that it should accurately recover the ejected mass of most input SNe Ia from
a range of contemporary progenitor scenarios.  We refrain from fine-tuning
our priors to match this suite of models, since it is a small set using one
radiation transfer code and any tuning attempts may be prone to overfitting,
but we explore some different plausible priors in order to bound the
associated systematics.


\subsection{Comparing Model Light Curve with Data}
\label{subsec:bololc-comp}

\begin{figure*}
\center
\resizebox{\textwidth}{!}{\includegraphics{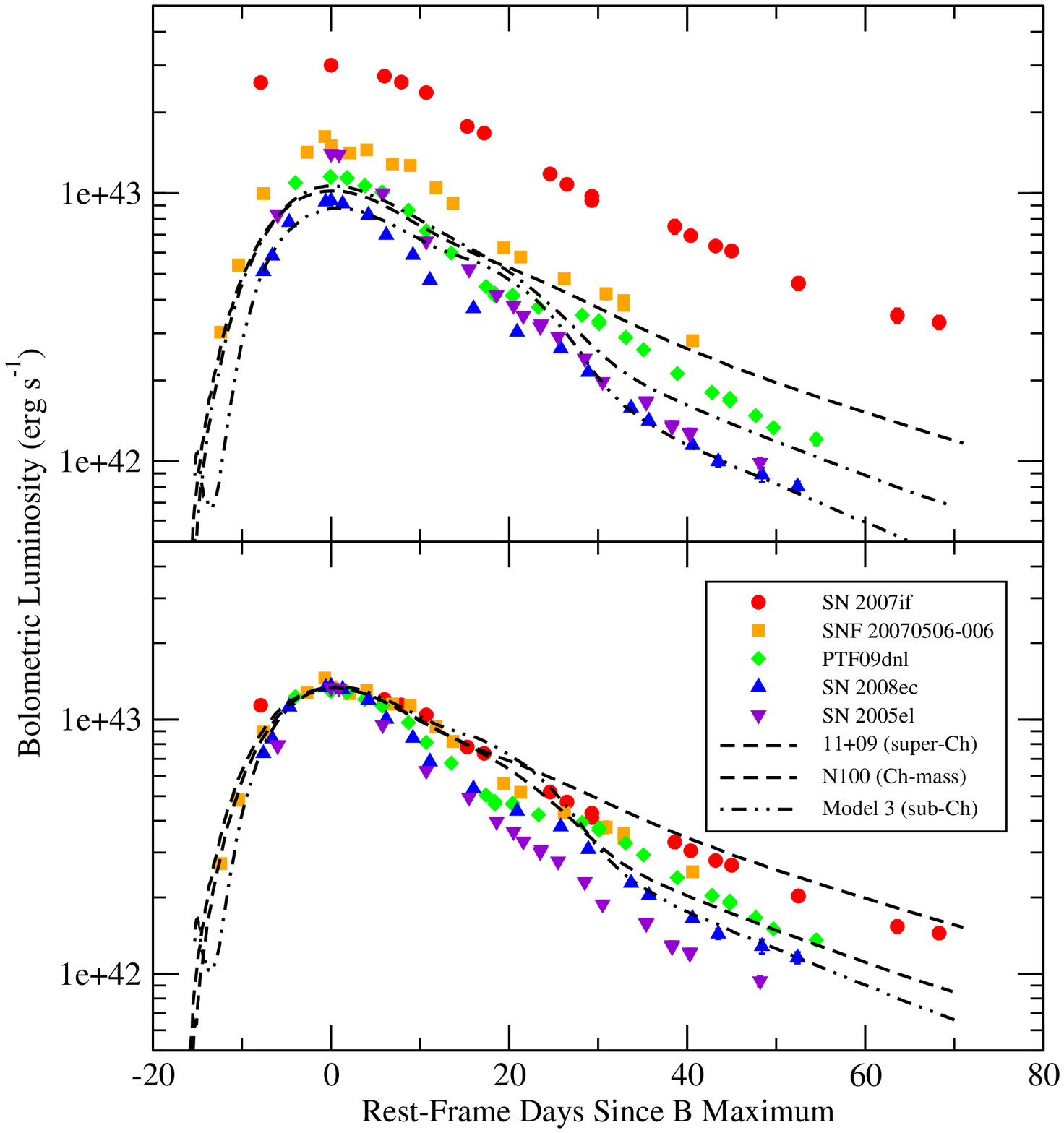}}
\caption{\small Observed bolometric light curves of representative SNfactory
SNe~Ia (coloured symbols with error bars), alongside synthetic observables
for explosion models (black curves).
Top:  original light curves; bottom: light curves normalized to a peak
luminosity of $1.2 \times 10^{43}$~erg~s$^{-1}$.}
\label{fig:bololc-comp}
\end{figure*}

To build confidence that our method is capturing useful distinctions between
SNe of different masses, we show a direct comparison between SNfactory light
curves and three representative explosion models in
Figure~\ref{fig:bololc-comp}.  The light curves as actually observed are
shown on the top, while on the bottom, they are
normalized to the same peak luminosity to emphasize differences in shape.

The models, all with $\MNi = 0.6$~\Msol\ but with differing ejected masses,
are shown as black curves.  The overall trend with light curve shape is clear:
the (angle-averaged) light curve of the super-Chandrasekhar-mass violent
merger 11+09 is the brightest at $+40$~days, followed by those of the
Chandrasekhar-mass delayed detonation N100 and the sub-Chandrasekhar-mass
double detonation Model~3.  The less massive models show inflections
corresponding to the NIR second maximum, but all have settled down into an
optically thin, quasi-exponential decline by $+40$ days \citep{jeffery99}.

The real SNfactory SNe span a broader range of \nickel\ mass and so show
a spread of absolute magnitudes, but the light curve shapes are usually
quite similar to the models for corresponding reconstructed masses.
SN~2007if, with $\MWD = 2.30^{+0.27}_{-0.24}$~\Msol\ \citep{scalzo12},
has a broad, uninflected light curve with a decay rate similar to
the 1.95-\Msol\ model 11+09; it is three times more luminous overall,
and seems to decline slightly more rapidly than 11+09.  The difference in
decline rate may be a sign that more radiation is being trapped or produced
near maximum light, or that more radiation is escaping at late times from
\cobalt\ in higher-velocity ejecta.
PTF09dnl ($\MWD = 1.33^{+0.13}_{-0.13}~\Msol$) closely resembles the
Chandrasekhar-mass model N100, and
SN~2008ec ($\MWD = 1.02^{+0.10}_{-0.09}~\Msol$) closely resembles the
sub-Chandrasekhar-mass Model 3.

SN~2005el presents an interesting
outlier case which we shall discuss in more detail in the following section.
It has a late-time light curve similar to the sub-Chandrasekhar-mass models,
but is as bright near maximum light as the Chandrasekhar-mass models;
this implies a very high \nickel\ content which should result in a peculiar
spectrum, but in fact it appears spectroscopically normal.


\subsection{Trends with Decline Rate}
\label{subsec:x1-plots}


\begin{figure*}
\center
\resizebox{\textwidth}{!}{\includegraphics{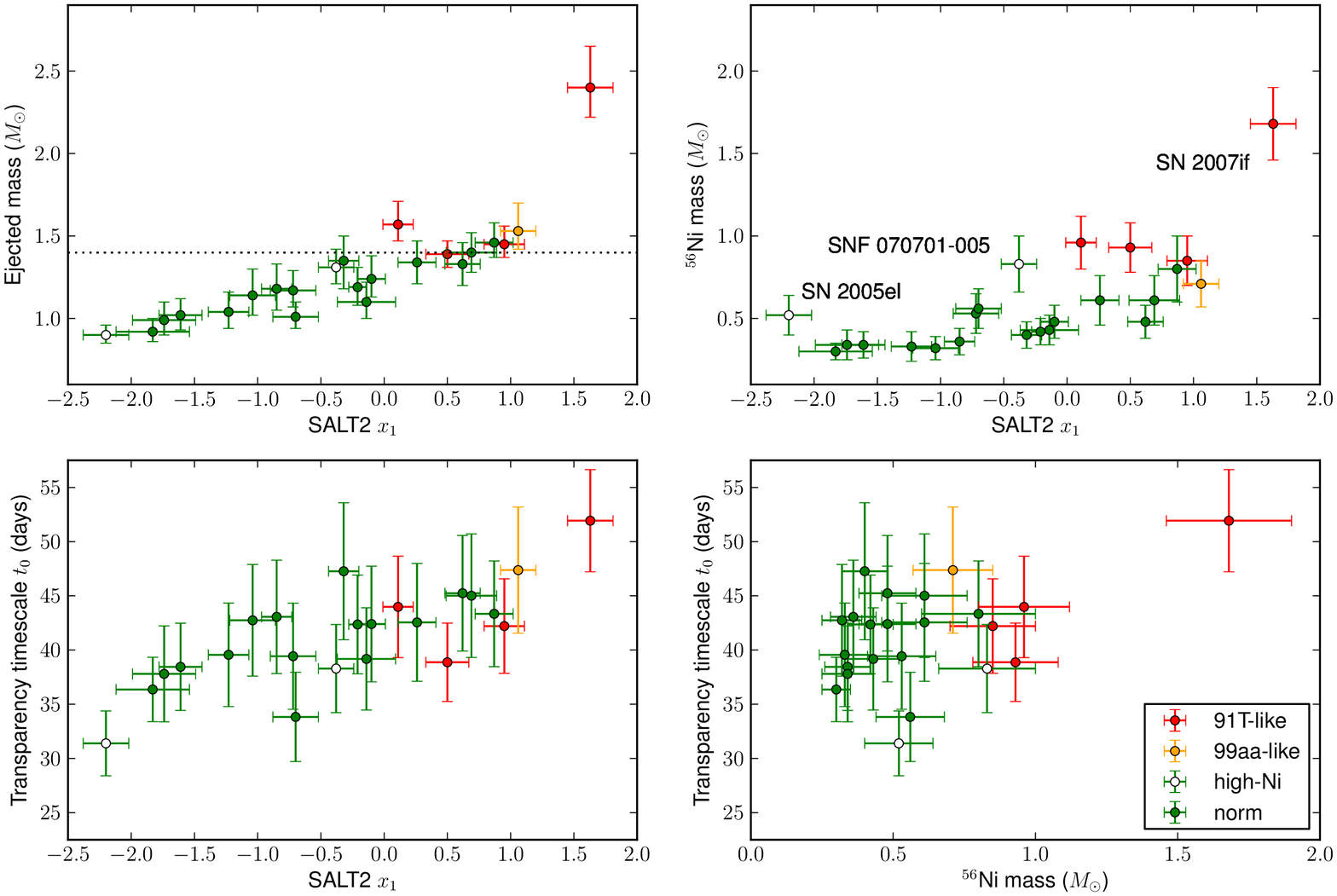}}
\caption{\small Correlations between reconstructed quantities and light curve
parameters.  Different colours show different spectroscopic subtypes:
red = 1991T-like/super-Chandra \citep{scalzo10,scalzo12},
orange = 1999aa-like, green = core normal.  Spectroscopically normal SNe~Ia
which show up as outliers in the \MWD-$t0$ plane are shown as open circles.
The horizontal dotted line marks the Chandrasekhar mass $M = 1.4~\Msol$.}
\label{fig:massplots-fiducial}
\end{figure*}

A correlation between light curve decline rate and ejected mass is expected
for SNe~Ia \citep[e.g.][]{arnett82}, and indeed for radioactively powered
SNe in general, since the diffusion time for optical photons should increase
with mass.  The scaling relations of \citet{arnett82} are frequently used by
observers to obtain rough estimates of the ejected masses of supernovae
\citep[e.g.][]{sullivan11a,drout11,cano13}.  However, the degeneracy between
the ejected mass and other factors affecting the diffusion time, including
the ejecta velocity and opacity to optical-wavelength photons, severely limits
the accuracy of mass predictions from near-maximum-light data.  Opacities in
particular depend on the temperature and composition and may therefore vary
with time \citep{kmh93}.  In contrast, our method, which relies on the
well-understood, nearly-gray opacity of Compton scattering in the
optically-thin limit \citep{swartz95,jeffery99}, has the potential to break
the degeneracies and shed light on the relationship between mass and
near-maximum-light decline rate.

Figure~\ref{fig:massplots-fiducial} shows the dependence of the underlying
parameters \MNi\ and $t_0$, and of the inferred mass \MWD, on the
light curve decline rate parameter $x_1$ for the SNfactory sample.  We have
colour-coded the points by spectroscopic subtype, showing 1991T-like and
1999aa-like SNe for comparison with the general population of normal SNe~Ia.

Most striking is the strength of the correlation between \MWD\ and $x_1$,
with very small dispersion.  A measurement of the light curve shape is
enough to determine the mass almost as accurately as the full fit.
A similar positive correlation is seen, as expected, in \MNi\ vs. $x_1$,
though with more variation.  Excluding two outliers which we shall discuss
below, the least-square best-fitting linear trends to the data for normal
SNe~Ia, taking both errors in $\MWD$ and $x_1$ into account, are
\begin{eqnarray}
\MWD/\Msol & = & (1.253 \pm 0.022) + (0.172 \pm 0.021)\,x_1 \\
\MNi/\Msol & = & (0.478 \pm 0.023) + (0.100 \pm 0.020)\,x_1
\end{eqnarray}
with Pearson's $r = 0.900$ ($p < 10^{-5}$) for \MWD\ vs. $x_1$. 
Although the true underlying trend may not in fact be linear, the reduced
chi-squares for both fits are small:  $\chi^2/\nu = 5.9/14 = 0.41$ for
a linear fit to \MWD\ vs. $x_1$, and $6.6/14 = 0.47$ for \MNi\ vs. $x_1$.
This suggests that some of the model-dependent
parameters over which we marginalize (such as $\alpha$) may be strongly
correlated with each other for a given SN, and/or may have similar values for
different SNe in our sample with similar $x_1$, although the true values of
these parameters are not accurately known.
\snf{070506-006}, the only 1999aa-like SN~Ia in the SNfactory sample with
sufficiently high data quality at late times to be considered here,
reconstructs with mass $\MWD = 1.53^{+0.17}_{-0.11}~\Msol$, on the high end
of our mass range for \revised{spectroscopically normal SNe~Ia} but not
definitely super-Chandrasekhar-mass.
Seven SNe in our fiducial analysis reconstruct as sub-Chandrasekhar at
greater than 95\% confidence, of which five have $x_1 < -1$.

We re-emphasize that the \MWD-$x_1$ correlation is not a spurious trend
arising solely from
any explicit dependence on $x_1$ in our analysis chain.  The trend changes
negligibly when the \citet{ganesh11} rise-time prior is replaced by a
simple Gaussian prior $t_{R,\mathrm{bol}} = 17 \pm 2$~days, or when the
$x_1$-dependent NIR correction is replaced by a mean correction.
The dependence must therefore already be imprinted on the shape of the
post-maximum optical light curves, as shown in Figure~\ref{fig:bololc-comp}.

The transparency time $t_0$ also has a strong correlation with $x_1$,
and since $t_0$ is derived directly from the data, this correlation is harder
to explain as an artifact of our fitting procedure.  \cite{stritz06} noted
\revised{a} similar correlation using a much simpler set of priors.
We have also verified that we get the same results for two very well-sampled
light curves with different reconstructed masses, SN~2007if (super-Chandra)
and SN~2011fe (Chandrasekhar-mass), by fitting subsamples of the late-time
light curve data, first using a single point near $B$-band phase $+40$~days
and then again using only points later than $+60$~days.
The median reconstructed mass changes by less than 0.03~\Msol\ in each case.

Starting from the fast-declining end, $t_0$ increases sharply with $x_1$
at first; the slope decreases for $x_1 > -1$.  Such a break may also appear
in the \MWD-$x_1$ plane, although if it does, it is less dramatic.  Finally,
the plot of \MNi\ vs. $t_0$, the closest we can come to the raw data, shows
no particularly strong trend, although this is not in itself surprising since
the two parameters are functionally independent in the \citet{arnett82}
formalism.  SNe with 1991T-like and 1999aa-like maximum-light
spectra cluster at the slowly-declining, slowly-diffusing, high-\MNi,
high-\MWD\ end of each plot; all of the spectroscopically peculiar SNe~Ia
studied in \citet{scalzo12} have $\MNi > 0.8~\Msol$.

Two of our SNe, SN~2005el and \snf{070701-005}, are outliers in the
\MNi-$x_1$ plane.  The reconstructions for these two SNe
show very high \MNi\ ($\sim 0.7\,\MWD$) typical of 1991T-like or
super-Chandra SNe, yet they appear spectroscopically normal.
We discuss these below.

\snf{070701-005} originally reconstructs with
$\MWD = 1.31^{+0.11}_{-0.10}$~\Msol\ and $\MNi = 0.83 \pm 0.17$~\Msol.
The derived host galaxy reddening from \ewna\ and from the Lira relation are
nearly identical, making the intrinsic $B-V$ colour of the SN at $B$-band
maximum light near zero.  Our measured absolute magnitude for this SN is also
comparable to the 1999aa-like \snf{070506-006}, mentioned above.
The behavior of this SN near maximum light is less well-constrained than
for our other SNe, and the uncertainty on the reddening is larger, leading
to a larger uncertainty in the \nickel\ mass.  This SN may simply have
scattered up on the diagram, or may show mild departures from the particular
assumptions of our method.
We expect our ejected mass estimate to be relatively robust to large
uncertainties in the \nickel\ mass (see \S\ref{subsec:xchckes});
note also \snf{080717-000}, which has the most uncertain \nickel\ mass
estimate in our sample, but for which the ejected mass is relatively
well-constrained.  The ejected mass of \snf{070701-005} is consistent with
the Chandrasekhar mass and its behavior is not unusual in any other respect.

SN~2005el presents a more interesting case.  It has one of the best-sampled
SNIFS spectrophotometric time series, with several late-time points and
reproducible bolometric light curve precision at the 0.02~mag level.
It is the fastest-declining SN in our sample ($x_1 = -2.20$), with a robust
NIR second maximum, \ewna\ and Lira excesses consistent with zero reddening,
and a peak absolute bolometric luminosity of $1.3 \times 10^{43}$ erg~s$^{-1}$,
consistent with $\sim 0.6~\Msol$ of \nickel\ under Arnett's rule.
Others have confirmed these observed properties \citep{phillips07,hicken09}.
SN~2005el may have physical properties which are not
well-represented by our model.  The least exotic possibility is that our
priors are wrong, and that this SN is best described with a higher value of
$\alpha$ and/or a shorter rise time, so that less \nickel\ is required to
describe the peak bolometric luminosity we measure.  The value of $\alpha$
required to make SN~2005el resemble \snf{080918-004}, which has the most
similar mass, must be very large, at least 1.6.  SN~2005el could also have
an unusual density structure, or could be asymmetric.  In any case, if our
mass reconstruction is correct, it is more likely that SN~2005el actually
has less \nickel\ than our fiducial analysis suggests.


\subsection{Trends with Color and $EW(\mathrm{Na~I~D})$}
\label{subsec:color-plots}


\begin{figure*}
\center
\resizebox{\textwidth}{!}{\includegraphics{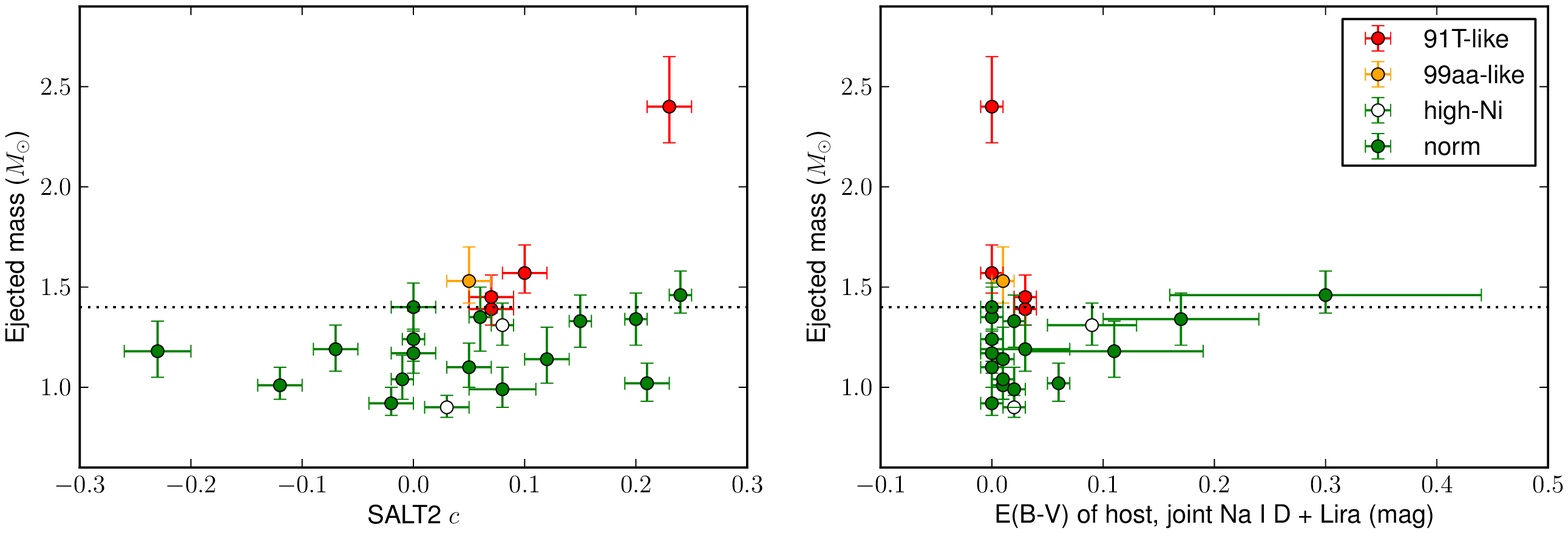}}
\caption{\small Correlations between reconstructed quantities and extinction
measures.  Colors represent different spectroscopic subtypes,
as in Figure~\ref{fig:massplots-fiducial}.
The horizontal dotted line marks the Chandrasekhar mass $M = 1.4~\Msol$.}
\label{fig:massplots-extdiag}
\end{figure*}

We are fortunate that most of the SNe in our sample show little or no
evidence for host galaxy reddening.  It is nevertheless worth checking
to see whether a trend with SALT2 $c$ or \ewna\ is apparent in the data.

Figure~\ref{fig:massplots-extdiag} shows the variation of \MNi\ and \MWD\
with SALT2 $c$ and with \ewna.  No obvious correlations appear.
Most of the SNe lie down at low \ewna, where a wide range of \MWD\
is seen.  Only three points have
\mbox{$E(B-V)_\mathrm{host} > 0.06$~\revised{mag}},
and these also have considerable uncertainty in the reddening.
SNe with large reddening corrections have uncertain \MNi\ and may plausibly
be biased towards higher \MWD.  However, our main conclusions ---
the existence of sub-Chandrasekhar-mass SNe~Ia, and of a correlation between
ejected mass and light curve width --- are not being driven by these SNe.


\subsection{Variation in Reconstruction Assumptions}
\label{subsec:xchckes}

\begin{table}
\center
\caption{Variations in priors for different reconstruction runs}
\begin{tabular}{lcccc}
\hline
   Run & $\rho(v)$\nb{a} & $Q$\nb{b} & $\alpha$ & $M_\mathrm{CO}/\MWD$ \\
\hline
A & exp    & std  & $1.2 \pm 0.2$ & $0.00 \pm 0.05$ \\
B & pow3x3 & std  & $1.2 \pm 0.2$ & $0.00 \pm 0.05$ \\
C & exp    & hole & $1.2 \pm 0.2$ & $0.00 \pm 0.05$ \\
D & pow3x3 & hole & $1.2 \pm 0.2$ & $0.00 \pm 0.05$ \\
E & exp    & std  & $1.0$         & $0.00 \pm 0.05$ \\
F & pow3x3 & std  & $1.0$         & $0.00 \pm 0.05$ \\
G & exp    & std  & $1.2 \pm 0.2$ & $< 1$           \\
H & pow3x3 & std  & $1.2 \pm 0.2$ & $< 1$           \\
\hline
S\nb{c} & exp & $2.0 \pm 0.6$ & $1.0$ & --- \\
\hline
\end{tabular}
\medskip \\
\flushleft
Quantities with error bars represent Gaussian priors for the reconstruction;
quantities with no error bars represent fixed parameters.
Fiducial priors:  $\rho(v) \sim \exp(-\sqrt{12}v/\vKE)$, $\alpha = 1.2 \pm 0.2$, no \nickel\ hole. \\
\nb{a}~{Density profile as a function of ejecta velocity: \\
        ``exp'' $\propto \exp(-\sqrt{12}v/\vKE)$, as in 1-D explosion models. \\
        ``pow3x3'' $\propto [1 + (v/\vKE)^{3}]^{-3}$, similar to 3-D models cited in this work.} \\
\nb{b}~{Variations in the assumed \nickel\ distribution, resulting
        in changes to the dependence of $Q$ on composition.  In ``std'',
        \nickel\ and (stable) Fe are mixed to form a central core underneath
        layers of partially burned material; in ``hole'', stable Fe is centrally
        concentrated due to neutronization, as in 1-D explosion models,
        displacing \nickel\ outwards.  In run G a fixed numerical value is used.} \\
\nb{c}~{Run reproducing the priors of \citet{stritz06}, which assumed $q = 0.33 \pm 0.10$
        (corresponding to our $Q = 2.0 \pm 0.6$), $\alpha = 1.0$, and exponential ejecta
        with $e$-folding velocity $v_e = 3000 \pm 300$~\kms\ ($\vKE = 10392 \pm 3118$~\kms).}

\label{tbl:massrecon-priors}
\end{table}

\begin{figure*}
\center
\resizebox{\textwidth}{!}{\includegraphics{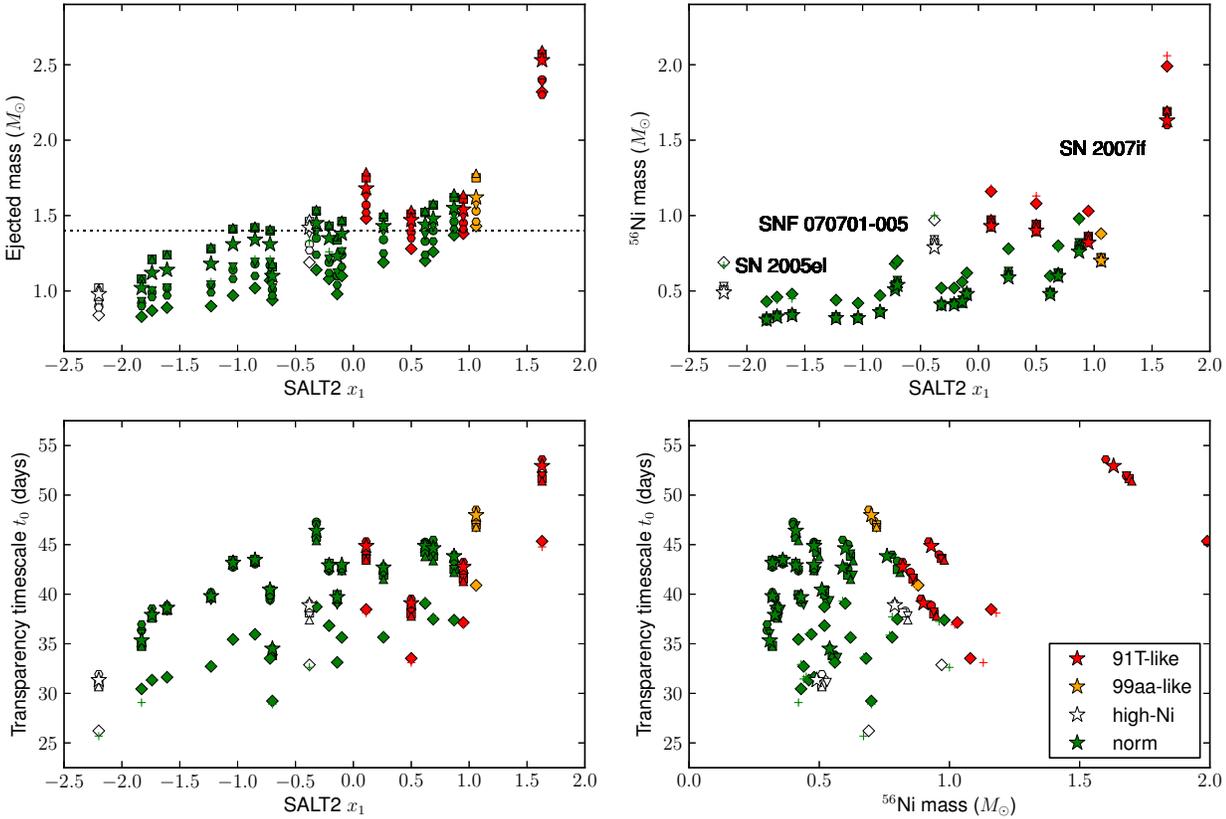}}
\caption{\small Influence of perturbations to input assumptions on
reconstructed quantities and their correlations with light curve parameters.
Colors represent different spectroscopic subtypes,
as in Figure~\ref{fig:massplots-fiducial}.  Error bars have been suppressed
to allow the mean values to be seen more clearly.
The horizontal dotted line marks the Chandrasekhar mass $M = 1.4~\Msol$.
Run A: circles; run B: squares; run C: inverted triangles; run D: triangles;
run E: diamonds; run F: crosses; run G: hexagons; run H: stars.
}
\label{fig:massplots-xchcke}
\end{figure*}

Although the priors for our fiducial analysis are well-motivated, they are
not unique, and performing many reconstructions with different input
assumptions can help quantify our sensitivity to these assumptions.
Some systematic effects, such as variations in $\alpha$, can be readily
parametrized and incorporated into our MCMC sampler, while others
(such as the radial dependence of the \nickel\ distribution) involve the
choice of a free function and/or lengthy calculations which are most
effective when decoupled from the MCMC.  We discuss such systematics
in this section.

\begin{table*}
\center
\caption{Ejected masses in different reconstruction runs}
\newcommand{\aatag}{\ensuremath{^\dagger}}
\begin{tabular}{lrrrrrrrrr}
\hline 
   SN~Name & Run A & Run B & Run C & Run D & Run E & Run F & Run G & Run H & Run S \\
\hline 
\multicolumn{10}{c}{SNfactory-Discovered Supernovae} \\
\hline 
SNF~20060907-000 & $1.01^{+0.09}_{-0.07}$ & $1.16^{+0.15}_{-0.11}$ & $1.02^{+0.10}_{-0.08}$ & $1.18^{+0.17}_{-0.12}$
                 & $0.94^{+0.05}_{-0.05}$ & $1.04^{+0.07}_{-0.06}$ & $0.97^{+0.09}_{-0.07}$ & $1.10^{+0.14}_{-0.11}$ & $0.99^{+0.49}_{-0.23}$ \\[0.2ex]
SNF~20061020-000 & $0.99^{+0.11}_{-0.09}$ & $1.21^{+0.18}_{-0.15}$ & $1.00^{+0.11}_{-0.09}$ & $1.21^{+0.18}_{-0.15}$
                 & $0.87^{+0.03}_{-0.03}$ & $1.01^{+0.06}_{-0.05}$ & $0.96^{+0.11}_{-0.08}$ & $1.12^{+0.19}_{-0.15}$ & $0.90^{+0.51}_{-0.26}$ \\[0.2ex]
SNF~20070506-006\aatag & $1.53^{+0.17}_{-0.11}$ & $1.75^{+0.32}_{-0.18}$ & $1.57^{+0.19}_{-0.13}$ & $1.78^{+0.30}_{-0.18}$
                 & $1.43^{+0.08}_{-0.06}$ & $1.58^{+0.13}_{-0.09}$ & $1.46^{+0.16}_{-0.12}$ & $1.62^{+0.28}_{-0.17}$ & $1.39^{+0.59}_{-0.34}$ \\[0.2ex]
SNF~20070701-005 & $1.31^{+0.11}_{-0.10}$ & $1.46^{+0.13}_{-0.10}$ & $1.38^{+0.14}_{-0.13}$ & $1.48^{+0.15}_{-0.11}$
                 & $1.19^{+0.06}_{-0.05}$ & $1.33^{+0.08}_{-0.07}$ & $1.27^{+0.12}_{-0.10}$ & $1.42^{+0.12}_{-0.12}$ & $1.44^{+0.50}_{-0.29}$ \\[0.2ex]
SNF~20070810-004 & $1.35^{+0.15}_{-0.17}$ & $1.53^{+0.25}_{-0.14}$ & $1.39^{+0.14}_{-0.18}$ & $1.54^{+0.25}_{-0.14}$
                 & $1.14^{+0.07}_{-0.06}$ & $1.36^{+0.11}_{-0.09}$ & $1.24^{+0.20}_{-0.17}$ & $1.45^{+0.22}_{-0.19}$ & $1.25^{+0.62}_{-0.39}$ \\[0.2ex]
SNF~20070817-003 & $1.04^{+0.12}_{-0.10}$ & $1.28^{+0.18}_{-0.17}$ & $1.04^{+0.13}_{-0.10}$ & $1.29^{+0.18}_{-0.17}$
                 & $0.90^{+0.04}_{-0.03}$ & $1.06^{+0.07}_{-0.06}$ & $0.99^{+0.12}_{-0.10}$ & $1.18^{+0.21}_{-0.17}$ & $0.97^{+0.55}_{-0.29}$ \\[0.2ex]
SNF~20070902-018 & $1.18^{+0.15}_{-0.13}$ & $1.42^{+0.18}_{-0.16}$ & $1.19^{+0.18}_{-0.13}$ & $1.43^{+0.17}_{-0.16}$
                 & $1.02^{+0.05}_{-0.04}$ & $1.21^{+0.08}_{-0.06}$ & $1.10^{+0.16}_{-0.13}$ & $1.34^{+0.19}_{-0.21}$ & $1.07^{+0.57}_{-0.32}$ \\[0.2ex]
SNF~20080522-011 & $1.40^{+0.12}_{-0.12}$ & $1.57^{+0.23}_{-0.14}$ & $1.43^{+0.13}_{-0.12}$ & $1.58^{+0.23}_{-0.14}$
                 & $1.26^{+0.07}_{-0.06}$ & $1.43^{+0.09}_{-0.06}$ & $1.33^{+0.15}_{-0.14}$ & $1.48^{+0.21}_{-0.14}$ & $1.26^{+0.55}_{-0.32}$ \\[0.2ex]
SNF~20080620-000 & $1.14^{+0.16}_{-0.12}$ & $1.41^{+0.17}_{-0.18}$ & $1.16^{+0.21}_{-0.14}$ & $1.42^{+0.17}_{-0.17}$
                 & $0.97^{+0.05}_{-0.05}$ & $1.16^{+0.08}_{-0.07}$ & $1.08^{+0.16}_{-0.13}$ & $1.31^{+0.19}_{-0.22}$ & $1.01^{+0.56}_{-0.32}$ \\[0.2ex]
SNF~20080717-000 & $1.46^{+0.12}_{-0.09}$ & $1.62^{+0.20}_{-0.14}$ & $1.50^{+0.15}_{-0.11}$ & $1.65^{+0.20}_{-0.15}$
                 & $1.37^{+0.08}_{-0.09}$ & $1.49^{+0.11}_{-0.08}$ & $1.41^{+0.12}_{-0.12}$ & $1.55^{+0.20}_{-0.14}$ & $1.55^{+0.55}_{-0.31}$ \\[0.2ex]
SNF~20080803-000 & $1.34^{+0.13}_{-0.13}$ & $1.49^{+0.18}_{-0.11}$ & $1.39^{+0.13}_{-0.15}$ & $1.51^{+0.19}_{-0.12}$
                 & $1.19^{+0.07}_{-0.07}$ & $1.36^{+0.09}_{-0.08}$ & $1.25^{+0.16}_{-0.13}$ & $1.43^{+0.17}_{-0.15}$ & $1.21^{+0.57}_{-0.31}$ \\[0.2ex]
SNF~20080913-031 & $1.10^{+0.12}_{-0.10}$ & $1.34^{+0.15}_{-0.17}$ & $1.12^{+0.16}_{-0.11}$ & $1.36^{+0.14}_{-0.17}$
                 & $0.98^{+0.04}_{-0.04}$ & $1.13^{+0.07}_{-0.06}$ & $1.04^{+0.12}_{-0.10}$ & $1.23^{+0.19}_{-0.17}$ & $0.99^{+0.53}_{-0.27}$ \\[0.2ex]
SNF~20080918-004 & $0.92^{+0.08}_{-0.06}$ & $1.08^{+0.14}_{-0.12}$ & $0.93^{+0.10}_{-0.07}$ & $1.08^{+0.14}_{-0.12}$
                 & $0.83^{+0.03}_{-0.02}$ & $0.90^{+0.05}_{-0.04}$ & $0.90^{+0.08}_{-0.06}$ & $1.02^{+0.14}_{-0.11}$ & $0.76^{+0.44}_{-0.21}$ \\[0.2ex]
\hline 
\multicolumn{10}{c}{Externally Discovered Supernovae Observed by SNfactory} \\
\hline 
SN~2005el        & $0.90^{+0.06}_{-0.05}$ & $1.02^{+0.10}_{-0.09}$ & $0.91^{+0.07}_{-0.05}$ & $1.02^{+0.10}_{-0.08}$
                 & $0.84^{+0.03}_{-0.02}$ & $0.90^{+0.04}_{-0.04}$ & $0.89^{+0.06}_{-0.05}$ & $0.98^{+0.10}_{-0.08}$ & $0.83^{+0.46}_{-0.20}$ \\[0.2ex]
SN~2007cq        & $1.17^{+0.12}_{-0.10}$ & $1.40^{+0.14}_{-0.15}$ & $1.19^{+0.16}_{-0.10}$ & $1.41^{+0.14}_{-0.15}$
                 & $1.07^{+0.05}_{-0.05}$ & $1.21^{+0.07}_{-0.06}$ & $1.11^{+0.12}_{-0.10}$ & $1.31^{+0.16}_{-0.17}$ & $1.05^{+0.53}_{-0.27}$ \\[0.2ex]
SN~2008ec        & $1.02^{+0.10}_{-0.09}$ & $1.24^{+0.16}_{-0.15}$ & $1.02^{+0.11}_{-0.09}$ & $1.24^{+0.17}_{-0.15}$
                 & $0.89^{+0.04}_{-0.04}$ & $1.02^{+0.05}_{-0.05}$ & $0.97^{+0.10}_{-0.09}$ & $1.14^{+0.18}_{-0.15}$ & $0.86^{+0.48}_{-0.25}$ \\[0.2ex]
SN~2011fe        & $1.19^{+0.12}_{-0.11}$ & $1.43^{+0.15}_{-0.13}$ & $1.21^{+0.14}_{-0.11}$ & $1.44^{+0.15}_{-0.13}$
                 & $1.08^{+0.06}_{-0.05}$ & $1.26^{+0.08}_{-0.07}$ & $1.12^{+0.14}_{-0.12}$ & $1.35^{+0.16}_{-0.20}$ & $1.12^{+0.58}_{-0.34}$ \\[0.2ex]
PTF09dlc         & $1.24^{+0.14}_{-0.11}$ & $1.46^{+0.17}_{-0.13}$ & $1.26^{+0.16}_{-0.12}$ & $1.47^{+0.17}_{-0.13}$
                 & $1.10^{+0.04}_{-0.04}$ & $1.27^{+0.08}_{-0.06}$ & $1.16^{+0.15}_{-0.12}$ & $1.38^{+0.16}_{-0.19}$ & $1.12^{+0.58}_{-0.31}$ \\[0.2ex]
PTF09dnl         & $1.33^{+0.13}_{-0.13}$ & $1.52^{+0.20}_{-0.12}$ & $1.37^{+0.13}_{-0.14}$ & $1.53^{+0.20}_{-0.12}$
                 & $1.20^{+0.06}_{-0.05}$ & $1.40^{+0.09}_{-0.07}$ & $1.24^{+0.17}_{-0.14}$ & $1.44^{+0.18}_{-0.17}$ & $1.45^{+0.63}_{-0.43}$ \\[0.2ex]
\hline 
\multicolumn{10}{c}{Numerical Explosion Models} \\
\hline 
Model 3          & $1.01^{+0.09}_{-0.08}$ & $1.22^{+0.15}_{-0.13}$ & $1.01^{+0.09}_{-0.08}$ & $1.23^{+0.16}_{-0.13}$
                 & $0.90^{+0.03}_{-0.03}$ & $1.06^{+0.06}_{-0.05}$ & $0.96^{+0.09}_{-0.08}$ & $1.14^{+0.16}_{-0.14}$ & $1.25^{+0.60}_{-0.39}$ \\[0.2ex]
Det\_1.10        & $1.22^{+0.12}_{-0.11}$ & $1.45^{+0.17}_{-0.12}$ & $1.24^{+0.16}_{-0.11}$ & $1.46^{+0.16}_{-0.11}$
                 & $1.04^{+0.04}_{-0.03}$ & $1.22^{+0.07}_{-0.06}$ & $1.13^{+0.14}_{-0.13}$ & $1.39^{+0.15}_{-0.21}$ & $1.45^{+0.63}_{-0.44}$ \\[0.2ex]
N5               & $1.35^{+0.11}_{-0.11}$ & $1.51^{+0.17}_{-0.11}$ & $1.39^{+0.11}_{-0.12}$ & $1.54^{+0.17}_{-0.12}$
                 & $1.20^{+0.05}_{-0.04}$ & $1.37^{+0.07}_{-0.07}$ & $1.27^{+0.14}_{-0.11}$ & $1.45^{+0.16}_{-0.13}$ & $1.41^{+0.51}_{-0.35}$ \\[0.2ex]
N100             & $1.27^{+0.14}_{-0.12}$ & $1.49^{+0.19}_{-0.11}$ & $1.29^{+0.15}_{-0.13}$ & $1.49^{+0.18}_{-0.11}$
                 & $1.11^{+0.05}_{-0.04}$ & $1.30^{+0.10}_{-0.07}$ & $1.18^{+0.17}_{-0.14}$ & $1.41^{+0.16}_{-0.19}$ & $1.57^{+0.62}_{-0.46}$ \\[0.2ex]
N1600            & $1.46^{+0.18}_{-0.11}$ & $1.72^{+0.39}_{-0.19}$ & $1.46^{+0.18}_{-0.11}$ & $1.73^{+0.39}_{-0.19}$
                 & $1.17^{+0.07}_{-0.05}$ & $1.43^{+0.15}_{-0.07}$ & $1.39^{+0.16}_{-0.21}$ & $1.59^{+0.36}_{-0.19}$ & $1.96^{+0.52}_{-0.51}$ \\[0.2ex]
11+09[1]         & $1.87^{+0.38}_{-0.18}$ & $2.25^{+0.59}_{-0.28}$ & $1.91^{+0.36}_{-0.19}$ & $2.28^{+0.57}_{-0.28}$
                 & $1.68^{+0.22}_{-0.10}$ & $2.01^{+0.46}_{-0.19}$ & $1.69^{+0.34}_{-0.23}$ & $1.98^{+0.60}_{-0.37}$ & $2.38^{+0.30}_{-0.44}$ \\[0.2ex]
11+09[2]         & $1.66^{+0.15}_{-0.10}$ & $1.88^{+0.26}_{-0.15}$ & $1.72^{+0.18}_{-0.12}$ & $1.92^{+0.25}_{-0.16}$
                 & $1.54^{+0.09}_{-0.06}$ & $1.70^{+0.13}_{-0.08}$ & $1.59^{+0.15}_{-0.11}$ & $1.77^{+0.24}_{-0.16}$ & $2.32^{+0.31}_{-0.55}$\\[0.2ex]
11+09[3]         & $1.59^{+0.22}_{-0.13}$ & $1.94^{+0.50}_{-0.22}$ & $1.62^{+0.22}_{-0.13}$ & $1.95^{+0.49}_{-0.21}$
                 & $1.46^{+0.12}_{-0.06}$ & $1.74^{+0.32}_{-0.12}$ & $1.49^{+0.21}_{-0.15}$ & $1.73^{+0.45}_{-0.26}$ & $2.25^{+0.37}_{-0.48}$ \\[0.2ex]
\hline 
\multicolumn{10}{c}{Run Statistics} \\
\hline 
Bias\nb{a} ($\sigma$)               & $-0.52$ & $+1.04$ & $-0.32$ & $+1.16$ & $-3.76$ & $-0.18$ & $-1.08$ & $+0.27$ & $+0.30$ \\[0.2ex]
Spread\nb{b} ($\sigma$)             & $ 0.95$ & $ 0.92$ & $ 0.86$ & $ 0.99$ & $ 1.59$ & $ 1.08$ & $ 0.86$ & $ 0.66$ & $ 0.90$ \\[0.2ex]
68\% CL accuracy\nb{c}              & $  5/8$ & $  3/8$ & $  5/8$ & $  3/8$ & $  0/8$ & $  6/8$ & $  4/8$ & $  6/8$ & $  7/8$ \\[0.2ex]
Non-$M_\mathrm{Ch}$ accuracy\nb{d}  & $  4/5$ & $  3/5$ & $  4/5$ & $  3/5$ & $  4/5$ & $  5/5$ & $  3/5$ & $  2/5$ & $  3/5$ \\[0.2ex]
$N(< M_\mathrm{Ch})$\nb{e}          & $ 7/16$ & $ 1/16$ & $ 4/16$ & $ 1/16$ & $15/16$ & $ 9/16$ & $10/16$ & $ 2/16$ & $ 0/16$ \\[0.2ex]
$N(> M_\mathrm{Ch})$\nb{f}          & $ 0/16$ & $ 1/16$ & $ 0/16$ & $ 1/16$ & $ 0/16$ & $ 0/16$ & $ 0/16$ & $ 0/16$ & $ 0/16$ \\[0.2ex]
\hline 
\end{tabular}
\medskip \\
\flushleft
Ejected masses reconstructed under assumptions different from the fiducial analysis.
Quantities with error bars are marginalized over all independent parameters.  Uncertainties represent
the 68\% CL intervals for the projections of the multi-dimensional PDF of the analysis in question.
Run priors are described in Table \ref{tbl:massrecon-priors}.\\
\aatag~Typed by SNID as 1999aa-like from multiple pre-maximum spectra. \\
\nb{a}~Mean of the pull distribution, i.e., the error-normalized residuals,
of the median reconstructed mass from the true value for simulated light curves of
3-D explosion models; this should be near zero for an accurate reconstruction.\\
\nb{b}~Standard deviation of the pull distribution; this should be near 1 for
properly estimated uncertainties. \\
\nb{c}~Number of explosion models for which the true value of the ejected
mass lies within the 68\% confidence interval. \\
\nb{d}~Number of non-Chandrasekhar-mass explosion models correctly identified
at high confidence ($> 95\%$~CL). \\
\nb{e}~Number of real SNe~Ia identified as sub-Chandrasekhar-mass at $> 95\%$~CL.\\
\nb{f}~Number of real SNe~Ia identified as super-Chandrasekhar-mass at $> 95\%$~CL.\\

\label{tbl:massrecon-reruns}
\end{table*}

Table~\ref{tbl:massrecon-priors} describes variations in the priors for
re-runs of our mass reconstruction.  As discussed in \ref{subsec:assumptions},
we vary priors on $\rho(v)$, on $\alpha$, on the mass of unburned material
$M_\mathrm{CO}$, and on the effect of neutronization on the \nickel\
distribution in the ejecta (influencing the transparency of the ejecta
through the form factor $Q$).
Table~\ref{tbl:massrecon-reruns} shows a comparison of the reconstructed
mass results under these different runs.
Figure~\ref{fig:massplots-xchcke} presents the same comparison visually,
showing a version of Figure~\ref{fig:massplots-fiducial}
overlaid with the results of different re-runs.

Not all of these re-runs necessarily correspond to plausible physics;
they are mainly meant to illustrate the impact of different assumptions.
To summarize our expectations for the biases introduced by a given set of
priors and their impact on our conclusions, we include at the bottom of
Table~\ref{tbl:massrecon-reruns} some summary statistics:  the mean and
standard deviation of the pull distribution, i.e., the error-normalized
residuals of our reconstructions from the simulated light curves; the number
of explosion models for which the true mass lies within our 68\% CL interval;
and the number of sub-Chandrasekhar-mass and super-Chandrasekhar-mass SNe~Ia
inferred in the SNfactory data set.

{\bf Runs A, C, F:}
Run A is our fiducial run, and the run we used for first-pass blind
validation of our method.  We would argue that run C, which assumes
$\alpha = 1.2 \pm 0.2$, exponential ejecta, and a central \nickel\ hole
due to neutronization, is best tuned to match 1-D explosion models in the
literature \citep[e.g.,][]{kmh93,hk96,blondin13}.
Run F, with $\alpha = 1.0$, power-law ejecta and no central \nickel\ hole,
is best tuned to match the 3-D explosion models we use for comparison in
\S\ref{subsec:mpa-challenge}.  As it turns out, these three runs make very
similar predictions:  all perform well on the suite of simulated light
curves, and all make similar predictions for the SNfactory SNe~Ia,
including a significant fraction of sub-Chandrasekhar-mass reconstructions.

{\bf Runs B, D, F, H:}
The choice of density profile has a significant effect on the absolute
mass scale for our reconstructions.  The bulk ejecta of the ``pow3x3''
profile have a roughly uniform density profile for $v < \vKE$, making them
less centrally concentrated than the ``exp'' profile for a given \vKE,
and making $Q$ less sensitive to variations in composition.  As a result,
relative to the ``exp'' cases, the mass scale shifts upwards by about
0.2~\Msol\ for all of our SNe, and the uncertainties increase modestly.

{\bf Runs C, D:}
In composition structures without a central \nickel\ hole, the presence of
additional stable iron-peak material has a minimal effect on the overall
radial distribution of \nickel\ in the ejecta.  The presence of a central
\nickel\ hole slightly increases our systematic uncertainty in $Q$; a large
central \nickel\ hole will in general reduce the column density seen by
\cobalt-decay gamma rays, reducing $Q$ and requiring a larger mass to
reproduce a given light curve shape.  The overall effect is quite small,
however, probably because the effects of neutronization are limited for
explosions at low central density (especially sub-Chandrasekhar solutions).

{\bf Runs E, F:}
Fixing $\alpha = 1.0$ brings the derived \nickel\ masses for the simulated
light curves closer into line with the true values.  The error bars also
decrease significantly, showing that understanding of $\alpha$ is a
limiting factor in our method's accuracy:  uncertainty in $\alpha$ affects
the light curve shape directly.  Run E (exponential density profile)
underestimates the ejected mass, but run F (power-law density profile)
performs very well on the simulated light curves, again unsurprising since
this set of priors is tuned specifically for these models.  Six of the eight
models have true masses within the 68\% CL interval; the pull distribution
has mean $-0.18$ and standard deviation $1.08$; and all of the SNe are
correctly identified as sub-Chandrasekhar, Chandrasekhar, or
super-Chandrasekhar.  Notably, with this choice
a large number of the SNfactory SNe~Ia in run F (9/16) reconstruct as
sub-Chandrasekhar-mass, even with a power-law density profile.

{\bf Runs G, H:}
Allowing the amount of unburned carbon to float freely tends to
\emph{decrease} the inferred mass.  A larger fraction of unburned carbon
means less nuclear energy released in the explosion, leading to lower kinetic
energy, more dense ejecta and hence a higher gamma-ray optical depth at
\revised{late} times.  Furthermore, given the moderately stratified
composition of our model ejecta, the unburned material is added on the
outside, further increasing the gamma-ray optical depth.  The data do not
in general allow more than 30\% of the white dwarf's original mass to remain
unburned, but allowing this much can shift the median reconstructed mass
downwards by up to 0.1~\Msol\ for some \revised{SNe}.
The direct impact of adding a variable amount of additional Compton-thick,
\nickel-poor material in the high-velocity ejecta also increases the
uncertainty on the inferred mass substantially, making it difficult to
identify non-Chandrasekhar-mass progenitors while not usefully improving
the accuracy of the reconstruction.

{\bf Run S:}
We include a reconstruction of our SNe using the priors of \citet{stritz06}.
The results show the same correlation between ejected mass and decline rate
as we derived and as \citet{stritz06} noted.  Interestingly, the Stritzinger
model manages to successfully flag the three views of 11+09 as
super-Chandrasekhar-mass, but its large uncertainties miss the
sub-Chandrasekhar-mass models completely.  We take this to imply that the
simple Stritzinger priors are not far off the correct mean behavior,
but we believe that our technique is much more informative and allows us to
explore the parameter space of explosion models in more detail.

In summary, we find that different choices of priors can shift the zeropoint
of the \MWD-$x_1$ relation up or down within a full range of 0.2--0.3~\Msol,
changing the number of events we class as sub-Chandrasekhar-mass or
super-Chandrasekhar-mass at $> 95\%$~CL.  However, the significance and
slope of the \MWD-$x_1$ relation remain roughly the same in all cases.
Moreover, sub-Chandrasekhar-mass SNe~Ia appear in our data set for a variety
of plausible priors which others have used in the past.  For any set of priors
which allow us to successfully identify sub-Chandrasekhar-mass supernovae
in our test suite of simulated light curves, we also find
sub-Chandrasekhar-mass SNe~Ia in our data.

Our method assumes spherical symmetry, and in this sense represents the
angle-averaged version of potentially asymmetric SNe~Ia.  Although the net
effects of asymmetry are not entirely obvious, one effect we expect it to
have is to produce variations in the luminosity of the event, depending on
how \nickel\ is distributed in the ejecta with respect to the line of sight.
One might expect these effects to be lower for events with large \nickel\
mass fractions, since the \nickel\ will then be distributed more evenly
among viewing angles \citep[see e.g.][]{maeda11}, and most pronounced among
faint events.  However, to the extent that different lines of sight of an
asymmetric event produce similar light curve shapes, our ejected mass
estimates should be relatively insensitive to asymmetries.  This is borne out
by our method's performance on the highly asymmetric violent merger model
11+09.  Ongoing simulations of violent mergers and other asymmetric explosions
should help to determine the full implications of asymmetry for our results.

Finally, some of the variations in explosion physics we have examined may be
correlated in ways not captured by our models.  If this is the case, however,
our results can still provide interesting constraints on the allowed
parameter space for explosion models.  For example, if $\alpha$
strongly \emph{anti}-correlates with light curve width, this might allow our
semi-analytic light curves to reproduce fast-declining SNe with
Chandrasekhar-mass models.  This particular case seems physically very
unlikely in the context of the explosion models we cite herein:  the 1-D
explosion models of \citet{hk96} actually show a \emph{correlation} with
positive sign between $\alpha$ (labelled $Q$ in table~2 of that paper)
and light curve width (rise time), though with large scatter, and in
general we expect larger $\alpha$ to be associated with more extensive
radiation trapping and longer rise times in the context of 1-D models.
Such a case is nevertheless indicative of the kind of constraint on
Chandrasekhar-mass models our results represent.


\section{Discussion}
\label{sec:discussion}

Although many variables could in principle alter our reconstruction, and the
absolute mass scale of our reconstructions may still be uncertain at the
15\% level based on those systematic effects we have been able to quantify,
we believe we have convincingly demonstrated that a range of SN~Ia progenitor
masses must exist.  For those sets of assumptions that incur minimal bias
when reconstructing simulated light curves, we find a significant fraction
(up to 50\%) of sub-Chandrasekhar-mass SNe~Ia in our real data.  We should
therefore take seriously the possibility that SNe~Ia are dominated by a
channel which can accomodate sub-Chandrasekhar-mass progenitors, or that at
least two progenitor channels contribute significantly to the total rate
of normal SNe~Ia.  We now attempt to further constrain progenitor models
by examining the dependence of \MWD\ on \MNi, with the caveat that the
systematic errors on \MNi\ may be larger than our reconstruction estimates.

The most mature explosion models currently available in the literature for
sub-Chandrasekhar-mass white dwarfs leading to normal SNe~Ia are those of
\citet{fink10}, with radiation transfer computed by \citet{kromer10},
and those of \citet{wk11}.  According to \citet{fink10}, systems with total
masses (carbon-oxygen white dwarf plus helium layer) as low as 1~\Msol\
can still produce up to 0.34~\Msol\ of \nickel.  The mass fraction of
\nickel\ increases rapidly \revised{with progenitor mass},
with the detonation of a \revised{1.29~\Msol}\ system producing
1.05~\Msol\ of \nickel.  \citet{wk11} find a similar trend, with nickel
masses ranging from 0.3--0.9~\Msol\ for progenitors with masses in the
range 0.8--1.1~\Msol.  The models differ in their prescriptions for igniting
a carbon detonation and in the resulting nucleosynthesis from helium
burning, but the overall \nickel\ yields agree in cases where a carbon
detonation has been achieved.

Very recently, the possibility of collisions of white dwarfs producing
SNe~Ia has also been raised \citep{benz89,rosswog09,raskin09}.  Ordinarily one
would expect white dwarf collisions to occur only in very dense stellar
environments such as globular clusters.  However, in triple systems consisting
of two white dwarfs accompanied by a third star in a highly eccentric orbit,
Kozai resonances can substantially decrease the time to a double-degenerate
merger or collision \citep{kd12,kushnir13}.
Both sub-Chandrasekhar-mass and super-Chandrasekhar-mass
SNe~Ia could arise through this channel.  The uncertainties involved in
predicting the rate of such events are substantial, but \cite{kushnir13}
make a concrete prediction for the variation of \nickel\ mass with total
system mass in white dwarf collisions, which we can evaluate here.
We caution that \citet{raskin10} show that \nickel\ mass, and indeed
the very occurrence of an explosion, depend on the mass ratio as well as
the impact parameter for the collision.

\begin{figure}
\center
\resizebox{0.5\textwidth}{!}{\includegraphics{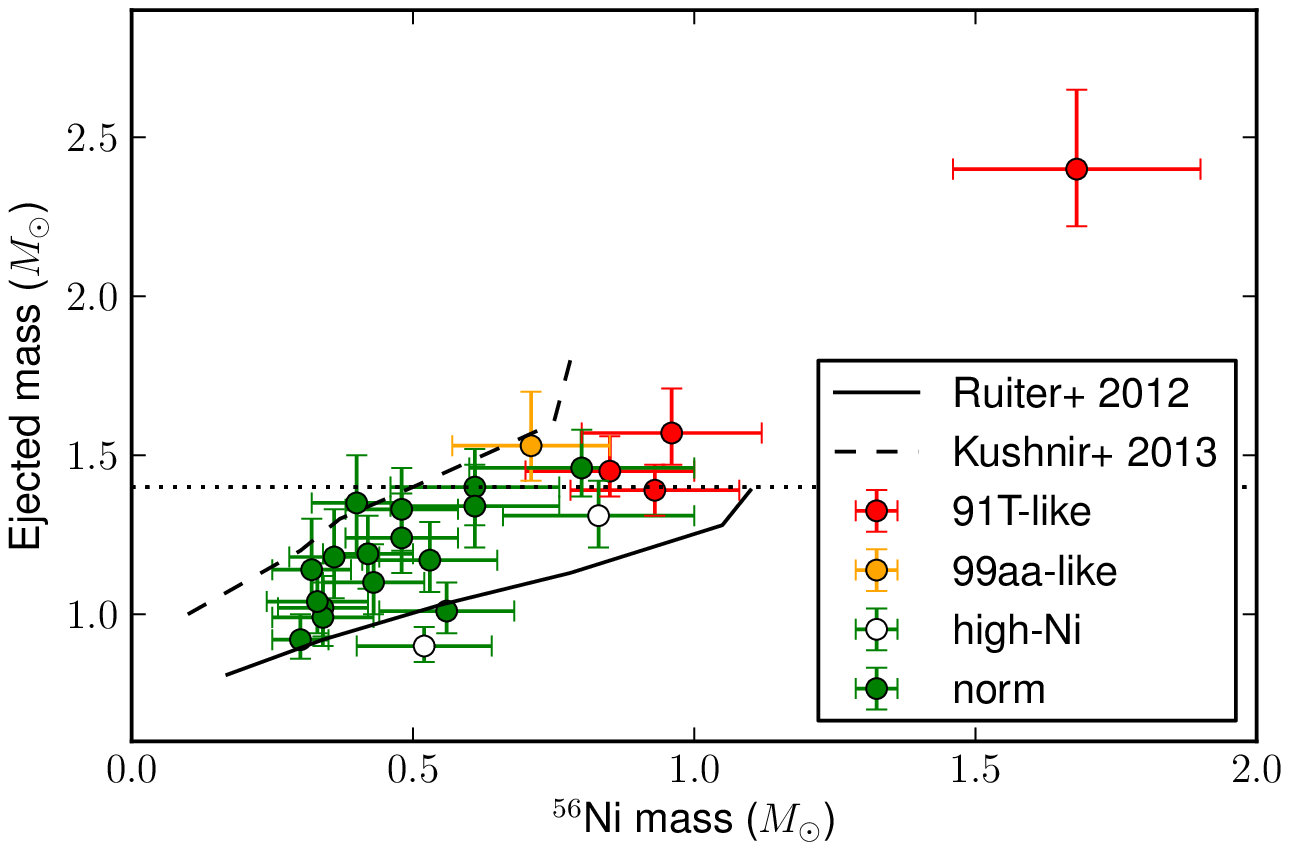}}
\caption{\small Ejected mass vs. \nickel\ mass for the SNfactory sample
in our fiducial analysis.  Colors represent different spectroscopic subtypes,
as in Figure~\ref{fig:massplots-fiducial}.
The horizontal dotted line marks the Chandrasekhar mass $M = 1.4~\Msol$.
The black solid curve shows the expected \MWD-\MNi\ relation for
sub-Chandrasekhar-mass double detonations from the models of \citet{fink10}
as presented in \citet{ruiter13}.  The dashed curve shows the predictions
of the white dwarf collision model of \citet{kushnir13}.}
\label{fig:massplots-mwdvmni}
\end{figure}

Figure~\ref{fig:massplots-mwdvmni} shows \MWD\ vs. \MNi\ for the SNfactory
data and the expected relations for the models of \citet{ruiter13} and
\citet{kushnir13}.  The \citet{ruiter13} trend seems to be consistent with
a few of the lowest-mass SNfactory SNe~Ia, but in general the predicted
increase of \MNi\ with \MWD\ is too steep to accommodate most of our
observations.  The trend of \citet{kushnir13} does reasonably well
for some of the low-\MNi\ SNfactory SNe~Ia, but can accommodate neither
our least massive SNe nor bright 1991T-like SNe~Ia.  The latter could perhaps
be explained by the more detailed collision models of \citet{raskin10}.

Interestingly, our SNe~Ia with $\MWD > 1.3~\Msol$ lie in a locus parallel
to the \citet{ruiter13} curve and about 0.3~\Msol\ higher.  While these
higher-mass SNe~Ia cannot easily be explained by double detonations, they
could perhaps be explained more naturally as double-degenerate mergers.
The violent merger models of \citet{pakmor10,pakmor11,pakmor12} are expected
to produce similar \nickel\ yields to double-detonation models with
comparable primary white dwarf masses \citep{ruiter13}.  Reproducing the
\nickel\ masses \revised{from our reconstruction} requires a primary white
dwarf mass of at least 1.1~\Msol.  However, \citet{pakmor11} showed that in
violent mergers of two carbon-oxygen white dwarfs, a mass ratio of at least
0.8 is needed to trigger the explosion, meaning that violent mergers with
$\MNi > 0.5~\Msol$ should have $\MWD > 1.9~\Msol$, like the different
views of 11+09 listed in Table~\ref{tbl:massrecon-mpa}
(which our method correctly reconstructed as super-Chandrasekhar).
Our absolute mass scale would have to be inaccurate \revised{at 50\% level}
to explain our observations with current models of violent mergers of
two carbon-oxygen white dwarfs.  The trend could also be generated by violent
mergers of a carbon-oxygen white dwarf with a helium white dwarf
\citep{pakmor13}, since helium ignites more readily than carbon and
a near-equal mass ratio is therefore not necessary.  More work is needed to
understand whether such mergers with system masses and synthesized \nickel\
masses consistent with our observations would appear spectroscopically normal.

The simplest explanation is that more massive, more \nickel-rich
SNe~\revised{Ia} are
Chandrasekhar-mass delayed detonations, arising either from slow mergers
of double-degenerate systems \citep{it84} or from single-degenerate systems.
Double-detonation models with
$\MWD > 1.15~\Msol$ have \MNi-\MWD\ ratios which should result in peculiar
spectra.  The mean mass of our normal SNe~Ia above this threshold
(of which there are eight) is $1.31 \pm 0.02~\Msol$ (stat), within 0.1~\Msol\
of the Chandrasekhar mass; this increases to $1.36 \pm 0.02~\Msol$
if the \citet{scalzo12} SNe (i.e., other than SN~2007if) are included.

Thus, according to our best current models, the data require at least two
progenitor scenarios:  one for sub-Chandrasekhar-mass white dwarfs, and one
or more for Chandrasekhar-mass and more massive white dwarfs, which could
arise from a variety of channels including Chandrasekhar-mass delayed
detonations, double-degenerate violent mergers, or possibly
spin-down single- or double-degenerate models resulting in a single
super-Chandrasekhar-mass white dwarf.  Since we have modeled only the
bolometric light curves, with no details of the spectroscopic evolution or
other observables (such as polarization or evidence for weak CSM interaction),
our results should not be taken to prescribe any particular subset of
explosion models of a particular mass.  However, any successful model
or suite of models should be able to reproduce our findings.


\section{Conclusions}
\label{sec:conclusions}

We have demonstrated a method to reconstruct the ejected masses of normal
SNe~Ia using Bayesian inference.  The method uses the semi-analytic
formalism of \citet{jeffery99} to compute the predicted late-time bolometric
light curve from \cobalt\ decay for a SN~Ia of a given ejected mass;
it is similar to the method of \citet{stritz06}, but includes more realistic
near-infrared corrections and more useful priors on unobserved variables.
Applying the method to a sample of SNfactory SNe~Ia with observations at
appropriately late phases, and to a suite of synthetic light curves from
full three-dimensional radiation transfer simulations of SNe~Ia,
we have shown the following:
\begin{enumerate}
\item The reconstructed ejecta mass is strongly correlated with the light
      curve width measured using cosmological light curve fitters,
      with a slope significantly different from zero.  We interpret this as
      strong evidence for a range of ejected masses in SNe~Ia.  Even if the
      range of masses is not as wide as our fiducial reconstruction suggests,
      due to variation in the density profiles or \nickel\ distributions
      which we do not directly constrain, any suite of explosion models
      intending to explain normal SNe~Ia must reproduce this correlation.
\item Our derived values for the ejected mass are relatively insensitive to
      systematic uncertainties in the \nickel\ mass, to mild asymmetry in
      the ejecta, and presumably to any systematic which does not affect
      the shape of the bolometric light curve.  The systematic error in our
      overall reconstructed mass scale associated with effects we are able
      to quantify is about $\pm 0.15$~\Msol.  This gives us further confidence
      that we are actually constraining the ejected masses of these SNe.
      Our most influential systematics are the unknown degree of radiation
      trapping near maximum light (parametrized by $\alpha$) and the
      influence of the ejecta density profile.
\item Ejected masses can be reconstructed via this method using a single
      observation of sufficiently high signal-to-noise at +40 days after
      bolometric maximum light, though with a mild bias towards low masses
      compared to a reconstruction done with a more complete light curve.
\item The observed locations of our mass estimates in the \MWD-\MNi\ plane
      are not all consistent with sub-Chandrasekhar-mass double-detonation
      models \citep{fink10,wk11,ruiter13}.  If these models are taken
      as representative of sub-Chandrasekhar-mass SN~Ia models in general,
      our results favor at least two progenitor channels for normal SNe~Ia.
\end{enumerate}

Although we have learned much from a fairly simple treatment of a fairly
small statistical sample of SNe~Ia, we should bear in mind the method's
limitations.  Semi-analytic treatments are necessarily approximate,
with their main advantage being speed.
They rely on simplified parametrizations of a number of complex
physical effects, and cannot predict the spectra of these events in detail,
so that spectroscopic information must be incorporated in a very schematic
way.  As numerical methods advance and large grids or libraries
of synthetic spectra from contemporary explosion models become available,
we may learn more by comparing spectra directly to the models
\citep[e.g.][]{blondin13,dessart13}.  In the meantime, however, some interplay
between semi-analytic and full numerical techniques may help us progress, with
the former incorporating useful prior information from the latter.

Our specific method assumes spherically symmetric ejecta and simplified
functional forms for the radial density profile and the \nickel\ distribution.
Although its performance on strongly asymmetric explosion models with
non-exponential density profiles is better at first glance than one might
expect, the impact of strong asymmetries or deviations from an exponential
density profile on our results are not yet understood in detail.
Extensions of the method that incorporate additional information to break
the degeneracy between viewing angle and colour or intrinsic brightness
\citep[along the lines of, e.g.][]{maeda11}, or which marginalize over
possible asymmetries, density perturbations, and \nickel\ distributions
to produce a more robust estimate of the systematic error, will help
us derive more accurate \nickel\ masses and ejected masses in the future.

Finally, a larger statistical sample is also highly desirable to replicate our
findings and to make further inferences about SN~Ia progenitor populations.
Applying our method to a larger sample of SNe~Ia with good late-time light
curves in different host galaxy environments
(including, potentially, highly extinguished SNe~Ia if NIR data are available
to constrain the extinction) should help us validate and calibrate
the relations between \MWD\ and $x_1$, and between \MWD\ and \MNi.
Use of these calibrated relations will then allow us to provide mass
measurements for a much larger sample of SNe, to determine the true
volumetric rates of SNe~Ia broken down by ejected mass and as a function of
redshift, and ultimately to compare to binary population synthesis models
for the progenitor channels of interest.  Knowledge of the progenitor mass
distribution for large samples of SNe~Ia used in future cosmological Hubble
diagrams should help to constrain the relative rates of possible progenitor
scenarios, thereby improving our understanding both of the dark energy and
of the tools we use to study it.


\section*{Acknowledgements}
\label{sec:acknowledgements}

The authors are grateful to the technical and scientific staffs of the
University of Hawaii 2.2-meter telescope, the W. M. Keck Observatory, Lick
Observatory, SOAR, and Palomar Observatory, to the QUEST-II collaboration,
and to HPWREN for their assistance in obtaining these data.
The authors wish to recognize and acknowledge the very significant cultural
role and reverence that the summit of Mauna Kea has always had within the
indigenous Hawaiian community.  We are most fortunate to have the opportunity
to conduct observations from this mountain.
This work was supported by the Director, Office of Science, Office of
High Energy Physics, of the U.S. Department of Energy under Contract No.
DE-AC02-05CH11231; by a grant from the Gordon \& Betty Moore Foundation;
and in France by support from CNRS/IN2P3, CNRS/INSU, and PNC.
Parts of this research were conducted by the Australian Research Council
Centre of Excellence for All-Sky Astrophysics (CAASTRO), through project
number CE110001020.
RS acknowledges support from ARC Laureate Grant FL0992131.
ST acknowledges support from the Transregional Collaborative Research Center 
TRR 33 ``The Dark Universe'' of the Deutsche Forschungs\-gemeinschaft.
YC acknowledges support from a Henri Chretien International Research
Grant administrated by the American Astronomical Society, and from the
France-Berkeley Fund.
NC acknowledges support from the Lyon
Institute of Origins under grant ANR-10-LABX-66.
This research used resources of the National Energy Research Scientific
Computing Center, which is supported by the Director, Office of Science,
Office of Advanced Scientific Computing Research, of the U.S. Department
of Energy under Contract No. DE-AC02-05CH11231.  We thank them for a generous
allocation of storage and computing time.
HPWREN is funded by National Science Foundation Grant Number ANI-0087344,
and the University of California, San Diego.
We thank Dan Birchall for his assistance in collecting data with SNIFS.


\appendix

\section{Gaussian Process Regression}
\label{sec:gp}

GP regression is a machine learning technique which can be used to fit smooth
curves to data.  Rather than specifying a fixed underlying functional form,
the curve itself is treated as a stochastic process, such that any two
points $x$, $x'$ on the curve have a joint Gaussian distribution described
by a covariance function $k(x,x';\Theta_i)$; the arguments $\Theta_i$ are a
set of hyperparameters which encode prior knowledge about the curve 
(for example, a correlation time-scale between consecutive light curve points)
in a Bayesian framework.  The hyperparameters can be trained
by maximum likelihood estimation, trading off complexity in the model with
the residuals of the data from the curve.  The process also generalizes to
multiple independent variables, or data ``features'', and the curves become
best-fitting hypersurfaces.

GP regression can be very useful in contexts where the underlying functional
form of a relation between data points is not known a priori, but is expected
to be smooth.  It is easier to apply than conventional Gaussian smoothing to
data which are unevenly sampled, such as light curves.  Moreover, a GP
regression fit can be viewed as a probability distribution in function space,
so that each draw from the fit corresponds to a possible realization of the
underlying trend which is consistent with the data and satisfies the
covariance function $k(x,x';\Theta_i)$ for the best-fit $\Theta_i$.
This property makes it straightforward to estimate errors on the range of
GP predictions at a given value of $x$ by Monte Carlo methods.

We use GP regression in several contexts in the analysis to follow,
implemented using the Python module \texttt{sklearn} \citep{sklearn}.


\subsection{Light Curve Fits}
\label{subsec:gp-lc}

For bolometric and single-band light curve fits, we use a squared-exponential
covariance function
$k(t,t') = e^{-0.5(t-t')^2/\tau^2} + \sigma^2 \delta(t-t')$,
with a single feature $t$ and two hyperparameters:  a correlation time-scale
$\tau$ in days, and a ``nugget'' term $\sigma$ describing the noise
(which we fix to be the median 1-$\sigma$ error in magnitudes).
While there is a slight variation in the correlation time-scale from SN
to SN, as might be expected, we find our data are well-represented
by GP fits with $0.5 < \tau < 2.0$, and fits outside this range generally
overfit the data or display pathological behavior; we therefore constrain
$\tau$ to lie in this range when fitting light curves.


\subsection{Near-Infrared Flux Corrections}
\label{subsec:gp-nir}

For the near-infrared corrections (see \S\ref{subsec:nircorr}),
we fit a GP with three parameters: rest-frame $B$-band phase $t$,
wavelength $\log\ \lambda$, and SALT2 $x_1$ (i.e., decline rate).
Near-infrared light curves show a characteristic second maximum occurring
between 25 and 35 days after $B$-band maximum light, the timing of which
correlates strongly with the $B$-band decline rate \citep{csp10}.
Slower-declining SNe~Ia have later-occurring NIR second maxima,
which can be understood in terms of a model in which the second maximum is
powered by the recombination of \ion{Fe}{3} to \ion{Fe}{2}, which
redistributes flux from bluer wavelengths into the NIR \citep{kasen06}.
Accounting for the dependence of the NIR behavior on decline rate can make
a difference of nearly 1~mag in $Y$ and $J$ at 40 days after $B$-band maximum
light.  We also allow for correlation between neighboring bands through the
wavelength parameter, with each band represented at its central wavelength.
While the $YJHK$ bands represent statistically independent measurements,
they have qualitatively similar behavior arising from a common physical
origin, and capturing the similarities in the GP fit can help improve
the statistical power of the GP prediction in each band.
The covariance function is
\begin{equation}
k({\bf x}, {\bf x}') =
   \exp\left[ ({\bf x}-{\bf x}')^T {\bf \Theta}
              ({\bf x}-{\bf x}')^T \right],
\end{equation}
where the feature vector is ${\bf x} = (t, x_1, \log(\lambda))$ and
the hyperparameters are
${\bf \Theta} = \mathrm{diag}(\Theta_t, \Theta_{x_1}, \Theta_\lambda)$.

Although the CSP data also show some variation in the \emph{contrast}
of the NIR second maximum, possibly correlating with different degrees of
mixing of \nickel\ in the outer layers of ejecta \citep{kasen06,csp10},
this behavior has little
influence on the NIR light curve after the second maximum.  We therefore
do not attempt to capture such variation here, since our modeling in
\S\ref{sec:modeling} requires accurate predictions only of the behavior at
maximum light (\nickel\ mass) and at phases after the NIR second maximum.



\begin{thebibliography}{199}

\footnotesize

\bibitem[\protect\citeauthoryear{Aldering \etal}{2002}]{snf}
Aldering, G., Adam, G., Antilogus, P., \etal\
   2002, Proc. SPIE, 4836, 61

\bibitem[\protect\citeauthoryear{Aldering \etal}{2006}]{snf2005gj}
Aldering, G., Antilogus, P., Bailey, S., \etal\
   2006, \apj, 650, 510

\bibitem[\protect\citeauthoryear{Arnett}{1982}]{arnett82}
Arnett, W. D.
   1982, \apj, 253, 785

\bibitem[\protect\citeauthoryear{Bacon \etal}{1995}]{bacon95}
Bacon, R., Adam, G., Baranne, A., \etal\
   1995, A\&AS, 113, 347 

\bibitem[\protect\citeauthoryear{Bacon \etal}{2000}]{bacon00}
Bacon, R., Emsellem, E., Copin, Y., \etal\ 2000,
   in ASP Conf. Ser. 195, \emph{Imaging the Universe in Three Dimensions},
   ed. W. van Breugel \& J. Bland-Hawthorn (San Francisco: ASP), 173 

\bibitem[\protect\citeauthoryear{Bacon \etal}{2001}]{bacon01}
Bacon, R., Copin, Y., Monnet, G., \etal\
   2001, MNRAS, 326, 23

\bibitem[\protect\citeauthoryear{Bailey \etal}{2009}]{sjb09}
Bailey, S., Aldering, G., Antilogus, P., \etal\
   2009, \aanda, 500, L17

\bibitem[\protect\citeauthoryear{Baltay \etal}{2007}]{baltay07}
Baltay, C., Rabinowitz, D., Andrews, P., \etal\
   2007, \pasp, 119, 1278


\bibitem[\protect\citeauthoryear{Benz \etal}{1989}]{benz89}
Benz, W., Thielemann, F.-K., \& Hills, J.~G.
   1989, \apj, 342, 986

\bibitem[\protect\citeauthoryear{Blondin \& Tonry}{2007}]{snid}
Blondin, S., \& Tonry, J.~L.
   2007, \apj, 666, 1024

\bibitem[\protect\citeauthoryear{Blondin \etal}{2011}]{blondin11}
Blondin, S., Kasen, D., Roepke, F., \etal\
   2011, \mnras, 417, 1280

\bibitem[\protect\citeauthoryear{Blondin \etal}{2013a}]{blondin13}
Blondin, S., Dessart, L., Hillier, D.~J., \etal\
   2013, \mnras, 429, 2127

\bibitem[\protect\citeauthoryear{Bloom \etal}{2012}]{bloom12}
Bloom, J.~S., Kasen, D., Shen, K.~J., \etal\
   2012, \apj, 744, L17

\bibitem[\protect\citeauthoryear{Bohlin \& Gilliland}{2004}]{bg04}
Bohlin, R.~C., \& Gilliland, R.~L.
   2004, AJ, 127, 3508

\bibitem[\protect\citeauthoryear{Bongard \etal}{2011}]{bongard11}
Bongard, S., Soulez, F., Thi{\'e}baut, {\'E}. \etal\
   2011, \mnras, 418, 258

\bibitem[\protect\citeauthoryear{Branch \etal}{1993}]{bfn93}
Branch, D., Fisher, A., \& Nugent, P.
   1993, \aj, 106, 2383

\bibitem[\protect\citeauthoryear{Buton \etal}{2013}]{buton13}
Buton, C., Copin, Y., Aldering, G., \etal\
   2013, \aanda, 549, A8

\bibitem[\protect\citeauthoryear{Cano \etal}{2013}]{cano13}
Cano, Z.
   2013, \mnras, 434, 1098

\bibitem[\protect\citeauthoryear{Cardelli \etal}{1988}]{cardelli}
Cardelli, J.~A., Clayton, G.~C. \& Mathis, J.~S.
   1988, \apj, 329, L33

\bibitem[\protect\citeauthoryear{Childress \etal}{2013}]{childress13}
Childress, M.~J., Aldering, G., Antilogus, P., \etal\
   2013, \apj, 770, 107

\bibitem[\protect\citeauthoryear{Dessart \etal}{2013}]{dessart13}
Dessart, L., Hillier, D.~J., Blondin, S., \etal\
   2013, \mnras, submitted (\arxiv{1308.6352})

\bibitem[\protect\citeauthoryear{Di~Stefano \& Kilic}{2012}]{rds12}
Di~Stefano, R. \& Kilic, M.
   2012, \apj, 759, 56

\bibitem[\protect\citeauthoryear{Drout \etal}{2011}]{drout11}
Drout, M.~R., Soderberg, A.~M., Gal-Yam, A., \etal\
   2011, \apj, 741, 97

\bibitem[\protect\citeauthoryear{Fink \etal}{2010}]{fink10}
Fink, M., R\"opke, F.~K., Hillebrandt, W., \etal
   2010, \aanda, 514, A53

\bibitem[\protect\citeauthoryear{Folatelli \etal}{2010}]{csp10}
Folatelli, G., Phillips, M.~M., Burns, C., R., \etal\
   2010, \aj, 139, 120

\bibitem[\protect\citeauthoryear{Folatelli \etal}{2011}]{folatelli11}
Folatelli, G., Phillips, M.~M., Morell, N., \etal\
   2011, \apj, 745, 74

\bibitem[\protect\citeauthoryear{Foley \& Kasen}{2010}]{fk10}
Foley, R.~J. \& Kasen, D.
   2010, \apj, 729, 55

\bibitem[\protect\citeauthoryear{Foreman-Mackey \etal}{2013}]{emcee}
Foreman-Mackey, D., \etal
   2013, \pasp, 125, 306

\bibitem[\protect\citeauthoryear{Ganeshalingam \etal}{2011}]{ganesh11}
Ganeshalingam, M., Li, W., \& Filippenko, A.~V.
   2011, MNRAS, 416, 2607

\bibitem[\protect\citeauthoryear{Goldhaber \etal}{2001}]{goldhaber01}
Goldhaber, G., Groom, D.~E., Kim, A.~G., \etal\
   2001, \apj, 558, 359

\bibitem[\protect\citeauthoryear{Guy \etal}{2007}]{guy07}
Guy, J., Astier, P., Baumont, S., \etal\
   2007, \aanda, 466, 11

\bibitem[\protect\citeauthoryear{Guy \etal}{2010}]{guy10}
Guy, J., Sullivan, M., Conley, A., \etal\
   2010, \aanda, 523, 7

\bibitem[\protect\citeauthoryear{Hachisu \etal}{2011}]{hachisu11}
Hachisu, I., Kato, M., Saio, H., \etal
   2012, \apj, 744, 69

\bibitem[\protect\citeauthoryear{Han \& Podsiadlowski}{2004}]{hp04}
Han, Z., \& Podsiadlowski, P.
   2004, \mnras, 350, 1301

\bibitem[\protect\citeauthoryear{Hicken \etal}{2009}]{hicken09}
Hicken, M., Wood-Vasey, M.~W., Blondin, S., \etal\
   2009, \apj, 700, 1097

\bibitem[\protect\citeauthoryear{H\"oflich \& Khohklov}{1996}]{hk96}
H\"oflich, P. \& Khohklov, A.
   1996, \apj, 457, 500

\bibitem[\protect\citeauthoryear{Howell \etal}{2005}]{superfit}
Howell, D.~A., Sullivan, M., Perrett, K., \etal\
   2005, \apj, 634, 1190

\bibitem[\protect\citeauthoryear{Howell \etal}{2006}]{howell06}
Howell, D.~A., Sullivan, M., Nugent, P.~E., \etal\
   2006, Nature, 443, 308

\bibitem[\protect\citeauthoryear{Howell \etal}{2009}]{howell09}
Howell, D.~A., Sullivan, M., Brown, E.~F., \etal\
   2009, \apj, 691, 661

\bibitem[\protect\citeauthoryear{Iben \& Tutukov}{1984}] {it84}
Iben, I. \& Tutukov, A.~V.
   1984, \apjs, 54, 335 

\bibitem[\protect\citeauthoryear{Jeffery}{1999}]{jeffery99}
Jeffery, D.~J.
   1999, \astroph{9907015}

\bibitem[\protect\citeauthoryear{Jeffery, Branch, \& Baron}{2006}]{jbb06}
Jeffery, D.~J., Branch, D., \& Baron, E.
   2006, \astroph{0609804}

\bibitem[\protect\citeauthoryear{Justham}{2011}]{justham11}
Justham, S.
   2011, \apj, 730, L34

\bibitem[\protect\citeauthoryear{Kasen}{2006}]{kasen06}
Kasen, D.
   2006, \apj, 649, 939

\bibitem[\protect\citeauthoryear{Katz \& Dong}{2012}]{kd12}
Katz, B. \& Dong, S.
   2012, \arxiv{1211.4584}

\bibitem[\protect\citeauthoryear{Kessler \etal}{2009}]{kessler09}
Kessler, R. \etal\
   2009, \apjs, 185, 32

\bibitem[\protect\citeauthoryear{Khokhlov \etal}{1993}]{kmh93}
Khokhlov, A., M\"uller, E. \& H\"oflich, P.
   1993, \aanda, 270, 223

\bibitem[\protect\citeauthoryear{Kromer \& Sim}{2009}]{ks09}
Kromer, M., \& Sim, S. A.
   2009, \mnras, 398, 1809

\bibitem[\protect\citeauthoryear{Kromer \etal}{2010}]{kromer10}
Kromer, M., Sim, S.~A., Fink, M., \etal
   2010, \apj, 719, 1067

\bibitem[\protect\citeauthoryear{Kromer \etal}{2013}]{kromer13}
Kromer, M., Fink, M., Stanishev, V., \etal
   2013, \mnras, 429, 2287

\bibitem[\protect\citeauthoryear{Krueger \etal}{2010}]{krueger10}
Krueger, B.~K. Jackson, A.~P., Townsley, D.~M., \etal
   2010, \apj, 719, L5

\bibitem[\protect\citeauthoryear{Krueger \etal}{2012}]{krueger12}
Krueger, B.~K. Jackson, A.~P., Townsley, D.~M., \etal
   2012, \apj, 757, 175

\bibitem[\protect\citeauthoryear{Kushnir \etal}{2013}]{kushnir13}
Kushnir, D., Katz, B., Dong, S., \etal\
   2013, \apj, 778, L37

\bibitem[\protect\citeauthoryear{Lantz \etal}{2004}]{snifs}
Lantz, B., Aldering, G., Antilogus, P., \etal\
   2004, Proc. SPIE, 5249, 146

\bibitem[\protect\citeauthoryear{Lepage}{1978}]{vegas}
Lepage, G.~P.
   1978, Journal of Computational Physics, 27, 192

\bibitem[\protect\citeauthoryear{Linder}{2006}]{linder06}
Linder, E. V.
   2006, \prd, 74, 103518

\bibitem[\protect\citeauthoryear{Madison \etal}{2005}]{sn2005el}
Madison, D.~R., Baek, M., Li, W., 2005, CBET 233, 1.

\bibitem[\protect\citeauthoryear{Maeda \& Iwamoto}{2009}]{mi09}
Maeda, K., \& Iwamoto, K.
   2009, \mnras, 394, 239

\bibitem[\protect\citeauthoryear{Maeda \etal}{2011}]{maeda11}
Maeda, K., Leloudas, S., Taubenberger, S., \etal\
   2011, \mnras, 413, 3075

\bibitem[\protect\citeauthoryear{Mazzali \etal}{2007}]{mazzali07}
Mazzali, P. A., Roepke, F., Benetti, S., \etal\
   2007, Science, 315, 825


\bibitem[\protect\citeauthoryear{Nomoto \etal}{1984}]{w7}
Nomoto, K., Thielemann, F.-K., \& Yokoi, K.
   1984, \apj, 286, 644

\bibitem[\protect\citeauthoryear{Nomoto \& Kondo}{1991}]{nk91}
Nomoto, K. \& Kondo, Y.
   1991, \apj, 367, L19 

\bibitem[\protect\citeauthoryear{Nugent \etal}{1995}]{nugent95}
Nugent, P., Branch, D., Baron, E., \etal\
   1995, \prl, 75, 394

\bibitem[\protect\citeauthoryear{Nugent \etal}{2009}]{ptf09dlc}
Nugent, P., Sullivan, M., Howell, D.~A., \etal
   2009, ATEL, 2174, 1.

\bibitem[\protect\citeauthoryear{Nugent \etal}{2011}]{sn2011fe}
Nugent,~P., Sullivan,~M., Howell, D.~A., \etal
   2011, CBET, 2792, 1.

\bibitem[\protect\citeauthoryear{Nugent \etal}{2011}]{nugent11}
Nugent, P., Sullivan, M., Cenko, S. B., \etal\
   2011, Nature, 480, 344

\bibitem[\protect\citeauthoryear{Orff \& Newton}{2007}]{sn2007cq}
T.~Orff, J.~Newton, CBET 983, 1 (2007).

\bibitem[\protect\citeauthoryear{Pakmor \etal}{2010}]{pakmor10}
Pakmor, R., Kromer, M., R\"opke, F.~K., \etal
   2010, Nature, 463, 61

\bibitem[\protect\citeauthoryear{Pakmor \etal}{2011}]{pakmor11}
Pakmor, R., Hachinger, S., R\"opke, F.~K., \etal
   2011, \aanda, 528, 117

\bibitem[\protect\citeauthoryear{Pakmor \etal}{2012}]{pakmor12}
Pakmor, R., Kromer, M., \& Taubenberger, S.
   2013, \apj, 747, L10

\bibitem[\protect\citeauthoryear{Pakmor \etal}{2013}]{pakmor13}
Pakmor, R., Kromer, M., Taubenberger, S., \etal
   2013, \apj, 770, L8

\bibitem[\protect\citeauthoryear{Pedregosa \etal}{2011}]{sklearn}
Pedregosa, F., Varoquaux, G., Gramfort, A., \etal
    2011, Journal of Machine Learning Research, 12, 2825

\bibitem[\protect\citeauthoryear{Perlmutter \etal}{1999}]{scp99}
Perlmutter, S., Aldering, G., Goldhaber, G., \etal\
   1999, \apj, 517, 565

\bibitem[\protect\citeauthoryear{Phillips \etal}{1999}]{phillips99}
Phillips, M.~M., Lira, P., Suntzeff, N.~B., \etal\
   1999, \aj, 118, 1766

\bibitem[\protect\citeauthoryear{Phillips \etal}{2007}]{phillips07}
Phillips, M.~M., Li, W., Frieman, J.~A., \etal\
   2007, \pasp, 119, 360

\bibitem[\protect\citeauthoryear{Poznanski \etal}{2012}]{ppb12}
Poznanski, D., Prochaska, J.~X., \& Bloom, J.~S.
   2012, \mnras, 426, 1465

\bibitem[\protect\citeauthoryear{Raskin \etal}{2009}]{raskin09}
Raskin, C., Timmes, F.~X., Scannapieco, E., \etal\
   2009, \mnras, 399, L156

\bibitem[\protect\citeauthoryear{Raskin \etal}{2010}]{raskin10}
Raskin, C., Scannapieco, E., Rockefeller, G., \etal\
   2010, \apj, 724, 111

\bibitem[\protect\citeauthoryear{Rasmussen \& Williams}{2006}]{rw06}
Rasmussen, C.~E. \& Williams, C.~K.~I.,
   2006, \emph{Gaussian Processes for Machine Learning}, MIT Press

\bibitem[\protect\citeauthoryear{Rex \etal}{2007}]{sn2008ec}
Rex, J., Li, W., Filippenko, A.~V., \etal\
   2008, CBET 1437, 1.

\bibitem[\protect\citeauthoryear{Riess \etal}{1996}]{riess96}
Riess, A.~G., Press, W.~H., \& Kirshner, R.~P.
   1996, \apj, 473, 88

\bibitem[\protect\citeauthoryear{Riess \etal}{1998}]{riess98}
Riess, A.~G., Filippenko, A.~V., Challis, P., \etal\
   1998, \aj, 116, 1009


\bibitem[\protect\citeauthoryear{Rosswog \etal}{2009}]{rosswog09}
Rosswog, S., Kasen, D., Guillochon, J., \etal\
   2009, \apj, 705, 128

\bibitem[\protect\citeauthoryear{Ruiter \etal}{2011}]{ruiter11}
Ruiter, A.~J., Belczynski, K., Sim, S.~A., \etal\
   2011, \mnras, 417, 408

\bibitem[\protect\citeauthoryear{Ruiter \etal}{2013}]{ruiter13}
Ruiter, A.~J., Sim, S.~A., Pakmor, R., \etal\
   2013, \mnras, 429, 1425

\bibitem[\protect\citeauthoryear{Scalzo \etal}{2010}]{scalzo10}
Scalzo, R.~A., Aldering, G., Antilogus, P., \etal\
   2010, \apj, 713, 1073

\bibitem[\protect\citeauthoryear{Scalzo \etal}{2012}]{scalzo12}
Scalzo, R.~A., Aldering, G., Antilogus, P., \etal\
   2012, \apj, 757, 12

\bibitem[\protect\citeauthoryear{Schlegel \etal}{1998}]{sfd}
Schlegel, D. J., Finkbeiner, D. P. \& Davis, M.
   1998, \apj, 500, 525

\bibitem[\protect\citeauthoryear{Seitenzahl \etal}{2009}]{seitenzahl09}
Seitenzahl, I., Meakin, C.~A., Townsley, D.~M., \etal\
   2009, \apj, 696, 515

\bibitem[\protect\citeauthoryear{Seitenzahl \etal}{2011}]{seitenzahl11}
Seitenzahl, I., Ciaraldi-Schoolmann, F., \& R\"opke, F.~K.
   2011, \mnras, 414, 2709

\bibitem[\protect\citeauthoryear{Seitenzahl \etal}{2013}]{seitenzahl13}
Seitenzahl, I., Ciaraldi-Schoolmann, F., R\"opke, F.~K., \etal\
   2013, \mnras, 429, 1156

\bibitem[\protect\citeauthoryear{Silverman \etal}{2011}]{silverman11}
Silverman, J. M. \etal\
   2011, \mnras, 410, 585

\bibitem[\protect\citeauthoryear{Sim \etal}{2010}]{sim10}
Sim, S.~A., R\"opke, F.~K., Hillebrandt, W., \etal\
   2010, \apj, 714, L52 

\bibitem[\protect\citeauthoryear{Sim \etal}{2012}]{sim12}
Sim, S.~A., Fink, M., Kromer, M., \etal\
   2012, \mnras, 420, 3003

\bibitem[\protect\citeauthoryear{Stritzinger \etal}{2006}]{stritz06}
Stritzinger, M., Leibundgut, B., Walch, S., \etal\
   2006, \aanda, 450, 241

\bibitem[\protect\citeauthoryear{Stritzinger \etal}{2011}]{csp11}
Stritzinger, M., Phillips, M. M., Boldt, L. N., \etal\
   2011, \aj, 142, 156

\bibitem[\protect\citeauthoryear{Sullivan \etal}{2011}]{sullivan11a}
Sullivan, M., Kasliwal, M.~M., Nugent, P.~E., \etal
   2011, \apj, 732, 118

\bibitem[\protect\citeauthoryear{Sullivan \etal}{2011}]{sullivan11b}
Sullivan, M., Guy, J., Conley, A., \etal
   2011, \apj, 737, 102

\bibitem[\protect\citeauthoryear{Suzuki \etal}{2012}]{suzuki12}
Suzuki, N., Rubin, D., Lidman, C., \etal
   2012, \apj, 746, 85

\bibitem[\protect\citeauthoryear{Swartz \etal}{1995}]{swartz95}
Swartz, D.~A., Sutherland, P.~G., \& Harkness, R.~P.
   1995, \apj, 446, 766

\bibitem[\protect\citeauthoryear{Taubenberger \etal}{2011}]{taub11}
Taubenberger, S., Benetti, S., Childress, M. \etal\
   2011, \mnras, 412, 2735

\bibitem[\protect\citeauthoryear{Thomas \etal}{2007}]{rct07}
Thomas, R.~C., Aldering, G., Antilogus, P., \etal\
   2007, \apj, 654, 53

\bibitem[\protect\citeauthoryear{Thomas \etal}{2011}]{rct11}
Thomas, R.~C., Aldering, G., Antilogus, P., \etal\
   2011, \apj, 743, 27

\bibitem[\protect\citeauthoryear{Tripp}{1998}]{tripp98}
Tripp, R.
   1998, \aanda, 331, 815

\bibitem[\protect\citeauthoryear{Turatto \etal}{2002}]{tbc02}
Turatto, M., Benetti, S., \& Cappellaro, E.
   2002, \astroph{0211219}

\bibitem[\protect\citeauthoryear{van Kerkwijk \etal}{2010}]{vkcj10}
van Kerkwijk, M., Chang, P., \& Justham, S.
   2010, \apj, 722, L157

\bibitem[\protect\citeauthoryear{Wang \etal}{2009}]{wang09}
Wang, X., Filippenko, A.~V., Ganeshalingam, M., \etal\
   2009, \apj, 699, L139

\bibitem[\protect\citeauthoryear{Wang \& Han}{2012}]{wh12}
Wang, B., \& Han, Z.
   2012, New Astronomy Reviews, 56, 122

\bibitem[\protect\citeauthoryear{Whelan \& Iben}{1973}]{wi73}
Whelan, J. \& Iben, I. J.
   1973, \apj, 186, 1007 

\bibitem[\protect\citeauthoryear{Woosley \& Weaver}{1994}]{ww94}
Woosley, S.~E. \& Weaver, T. A.
   1994, \apj, 423, 371

\bibitem[\protect\citeauthoryear{Woosley \& Kasen}{2011}]{wk11}
Woosley, S.~E. \& Kasen, D.
   2011, \apj, 734, 38

\bibitem[\protect\citeauthoryear{Yoon \& Langer}{2005}]{yl05}
Yoon, S.-C. \& Langer, N.
   2005, \aanda, 435, 967

\bibitem[\protect\citeauthoryear{Yuan \etal}{2010}]{yuan10}
Yuan, F., Quimby, R.~M., Wheeler, J.~C., \etal\
   2010, \apj, 715, 1338

\end{thebibliography}
\end{document}